\documentclass[useAMS,usenatbib]{mn2e}
\usepackage{graphics,psfig,epsfig,graphicx,rotating,url}

\title[Colours and stellar populations of galaxies]{Optical and near infra-red colours as a discriminant of the age and
metallicity of stellar populations}
\author[D. Carter et al.]{David Carter$^{1}$, 
Daniel J. B. Smith$^{1}$, Susan M. Percival$^1$, Ivan K. Baldry$^1$,\newauthor
Christopher A. Collins$^1$, Philip A. James$^1$, Maurizio Salaris$^1$,  
\newauthor Chris Simpson$^1$, John P. Stott$^1$, Bahram Mobasher$^2$
\\$^{1}$Astrophysics Research Institute, Liverpool John Moores 
University, Twelve Quays House, Egerton Wharf, Birkenhead CH41 1LD, UK.
\\$^{2}$University of California, Riverside, 900 University Ave., 
Riverside, CA 92521, USA.}
\begin{document}

\date{Accepted ........ Received........ }

\pagerange{\pageref{firstpage}--\pageref{lastpage}} \pubyear{2008}

\maketitle

\label{firstpage}

\begin{abstract}

We present a comprehensive analysis of the ability of current stellar
population models to reproduce the optical ($ugriz$) and near infra-red
($JHK$) colours of a small sample of well-studied nearby elliptical and
S0 galaxies. We find broad agreement between the ages and
metallicities derived using different population models, although
different models show different systematic deviations from the
measured broad-band fluxes. Although it is possible to constrain
Simple Stellar Population models to a well defined area in
age-metallicity space, there is a clear degeneracy between these
parameters even with such a full range of precise colours. 
The precision to which age and metallicity can be 
determined independently, using only broad band photometry 
with realistic errors, is 
$\Delta{[Fe/H]} \simeq 0.18$ and $\Delta{\log{Age}} \simeq 0.25$. 
To constrain the
populations and therefore the star formation history further it will
be necessary to combine broad-band optical-IR photometry with either
spectral line indices, or else photometry at wavelengths outside of
this range.
\end{abstract}

\begin{keywords}
galaxies: stellar content --- galaxies: elliptical and lenticular, 
cD
\end{keywords}

\section{Introduction}

Knowledge of the star formation history of galaxies is key to our
understanding of their formation and of environmental influences upon
their properties. However, the relationship between observable
quantities and the fundamental parameters of the stellar population is
complex. Observables are usually interpreted by comparison with
single-age, single-metallicity population (Simple Stellar Population
or SSP) models. An SSP model is derived from a set of stellar tracks
and isochrones, which are populated using an assumed Initial Mass
Function (IMF). Empirical or theoretical spectra are summed along the
isochrone, with appropriate weighting, producing the SSP spectral
energy distribution. These Isochrone Synthesis Models were refined by
Charlot \& Bruzual (1991), and Bruzual \& Charlot (1993) and found
broadly to reproduce both spectroscopic and photometric properties of
galaxies. However in detail the predictions are affected in a complex
way by the population age, metallicity, abundance ratios such as the
degree of enhancement of $\alpha$-elements relative to solar ratios,
and the IMF. Interpretation of gradients within individual galaxies,
or trends with global parameters such as galaxy luminosity amongst
samples, for instance in clusters, are subject to various degeneracies
of which the best known is that between age and metallicity for old
populations (Worthey 1994; Kodama \& Arimoto 1997).

%
%

Most attempts to break these degeneracies, and establish
luminosity-weighted stellar ages, metallicities and abundance ratios
have involved obtaining high quality spectra of individual galaxies. A
common technique is to compare plots of pairs of line indices with
model grids (e.g. Trager et al. 2000a,b, 2008; Kuntschner et al. 2001;
Poggianti et al. 2001a,b; Nelan et al. 2005; S\'anchez-Bl\'azquez et
al. 2006). Other authors define combinations of indices designed to
measure specific parameters (Vazdekis \& Arimoto 1999), or apply
Principal Component Analysis or other non-parametric techniques to
extract specific physical parameters from galaxy spectra (Panter et
al. 2003, 2007; Tojeiro et al. 2007; Ocvirk et al. 2006a,b; Koleva et
al. 2008). An alternative which represents a compromise between
analysis of high signal-to-noise spectra, which can necessarily be
applied only to limited samples, and broad band photometry, is the
technique of narrowband continuum photometry (Rakos et al.  2007,
2008).

These techniques are limited to bright galaxies, where spectra or
narrow band images of sufficiently high signal-to-noise ratio can be
obtained.  For fainter samples, of dwarf or high redshift galaxies, or
of star clusters associated with brighter galaxies, it is important to
establish how precisely stellar population parameters can be measured
from broad-band photometry alone. Optical broad-band photometry alone
cannot break the age-metallicity degeneracy (e.g. Kodama and Arimoto
1999), however extending the spectral range appears to provide
additional constraints. This has been demonstrated for optical - near
infra-red colours by several studies, starting with Bothun et
al. (1984) who argued that $J-K$ depends principally on metallicity,
while $B-H$ depends also on mean stellar age.  The same colour
combination was used used by Bothun \& Gregg (1990) in an
investigation of the star formation history of S0 galaxies.  Peletier,
Valentijn and Jameson (1990) used $U-B$, $B-V$ and $V-K$ colours;
Peletier \& Balcells (1996) $U-R$ vs. $R-K$; Bell \& de Jong (2000) $B-R$
vs $R-K$; and James et al. (2006) $B-K$ vs. $J-K$ in further studies of the
metallicities and star formation histories of unresolved stellar
populations in galaxies.  Some
success has also been found in the use of optical-ultraviolet colours,
with Kaviraj et al. (2007a,b) combining SDSS optical and GALEX
ultraviolet photometry in a study of the star formation histories of
low-redshift early-type galaxies.


%
%

One of the problems with all of these techniques is the
model-dependence of the results. Each of the studies discussed above
will use a specific set of stellar tracks and isochrones, and a
specific set of theoretical or empirical spectra. Differences between
the results of these studies are often the result of differences
between models rather than between observations. In this paper we use
a variety of popular SSP models to derive stellar population
parameters from accurate photometry in ($ugrizJHK$) passbands of a
sample of nearby E/S0 galaxies. By using well calibrated public
datasets (SDSS and 2MASS) and photometry in consistent, well-defined
apertures, the observational error is reduced and we can investigate
in detail the uncertainty caused by the modelling process, the
differences between popular sets of models, the difference between
scaled-solar and $\alpha$-enhanced models, and that caused by
different assumptions about the foreground extinction.  Moreover for
some galaxies our derived parameters can be compared directly with
spectroscopic studies (e.g. Trager et al. 2008).

A somewhat similar study has been carried out by Eminian et
al. (2008). These authors use public catalogue photometry in optical
and near-infrared bands for a sample drawn from the SDSS spectroscopic
catalog. Their study contains a high proportion of currently star
forming galaxies, and their focus is on the correlation between star
formation, dust, and near-IR colours, rather than the age and
metallicity of old populations. Their study relies on magnitudes from
two different catalogue sources with different effective apertures, so
while they use colours internal to each of the datasets (e.g. $g-r$ and
$Y-K$) they cannot use the SED over the entire range to constrain the
population. The much lower precision of their photometry of individual
galaxies is balanced by the much larger sample that they use.

Conroy et al. (2009) use a new set of SSP models, currently
unavailable to us, to model the broad band SDSS and 2MASS catalogue
magnitudes of samples of galaxies at low redshift and at redshift
$\sim$2. They emphasise the importance of understanding the Thermally
Pulsing Asymptotic Giant Branch (TP-AGB) phase of stellar evolution
and of the IMF, and examine the effect of both on broad-band
colours. They find that Horizontal Branch morphology impacts the
derived population parameters less than either the TP-AGB and the
IMF. They also conclude that the evolution of a multi-metallicity
population, particularly in passbands at wavelengths longer than $V$,
is equivalent to that of a single-metallicity population whose
metallicity is the mean of the multi-metallicity population. The study
of Conroy et al. (2009), as with that of Eminian et al. (2008), relies
on large samples of galaxies with individually much lower photometric
precision than that of our measurements.

Lee et al. (2007), in a study which focuses on the composite
populations found in spiral galaxies, find that because the youngest
population ($<$ 2 Gyr) dominates the light, broadband colours can partially 
break the age-metallicity degeneracy. However at older ages they find 
large scatter in the results obtained from models by different authors. 

A complementary approach is to apply similar techniques to
  extragalactic star clusters, in particular to the globular cluster
  populations of galaxies. This has the advantage that, although the
  cluster populations associated with galaxies can have multiple
  formation epochs and particularly metallicities, an individual
  cluster is more likely than a galaxy to contain stars of only a
  single age and metallicity. This is balanced against the
  disadvantage of using star clusters, that a comparatively small
  number of bright stars (on the AGB or on the upper main sequence,
  depending upon the age of the cluster) will contribute substantially
  to the luminosity, and that statistical fluctuations in these
  numbers will cause fluctuations in the photometry, and thus lead to
  additional uncertainty in the derived parameters of the stellar
  population. This is not a problem for the galaxies in our study,
  which will have typically $10^4$ times as many stars in our defined
  apertures.

Puzia et al. (2002), and Hempel \& Kissler-Patig (2004a) applied $V-I$
vs.  $V-K$ colour-colour plots to the study of unresolved stellar
populations in globular clusters, in order to distinguish the
metallicities of the separate cluster populations associated with the
parent galaxies. Hempel \& Kissler-Patig (2004b) add $U$ band
observations and show that the age resolution is enhanced, enabling
separate formation epochs to be distinguished as well. Kundu et al.
(2005) find a substantial intermediate age population associated with
the Virgo elliptical galaxy NGC4365, using $gIH$ photometry alone. 

Anders et al. (2004) examine the ability of broad band ($UBVRIJH$)
photometry to recover the age and metallicity of artificial clusters
generated with GALEV SSP models (Schulz et al. 2002). They set
observational errors to be 0.1 mag in all passbands in their study,
which is a somewhat larger value than we have in our observational
data. They find that the $U$ and $B$ bands are critical for determining
ages, and that the maximum available wavelength range is also
important.  Applying the same fitting technique to a sample of real
clusters, de Grijs et al. (2005) find that for young star clusters,
and now using HST filters and passbands, the estimates
for the cluster age have an approximately Gaussian distribution with
$\sigma_{log(t)} \leq 0.35$. They do not attempt to constrain the
metallicity of their clusters, but note that the effects of
uncertainties in metallicity and extinction on the age determined are
small. 

   Most of the Globular Cluster studies also rely on only one set of
  stellar population models, usually those of Bruzual \& Charlot
  (2003, hereafter BC03). However Hempel et al. (2005), in their
  appendix, compare the model predictions for ($V-I$) and ($V-H$) for
  BC03 models and those of Maraston (2005), and find that the
  predictions for these colours do not differ significantly except for
  metal poor populations with age $\le$ 2Gyr. Pessev et al. (2008)
present 2MASS-based JHK photometry of a sample of Magellanic Cloud
globular clusters, and combine this with $UBV$ photometry from the
literature to undertake a comparison of the performance of four sets
of stellar population models at distinguishing age and metallicity for
this sample. They find that all models reproduce the colours for old
clusters quite well, but that for young ages there are more
substantial differences.  Overall they conclude that the models of
BC03 give the best quantitative match to their data. This study is
carried out at lower metallicity than ours, but comparison with our
conclusions will be interesting nevertheless.

\section{Observational Data}
\label{sec:data}

Our sources of observational data are the image servers provided by
data release 6 of the Sloan Digital Sky Survey (SDSS; Adelman-McCarthy
et al.  2008) and by the 2 Micron All-Sky Survey (2MASS; Skrutskie et
al. 2006). We have chosen a sample of bright galaxies in the Coma
cluster, and a sample at intermediate luminosity from the Virgo
cluster to test our techniques. These samples are limited by practical
issues: for fainter galaxies in Coma the photon and read noise, which
are the dominant noise source in our 2MASS magnitudes, adversely
affect the precision, while for brighter galaxies in Virgo the galaxy
image is so large that it is impossible to determine sky levels on the
size of image provided by the SDSS image server. To sample a range of
absolute magnitude it was necessary to select galaxies from both
clusters.

SDSS and 2MASS images of a number of galaxies were downloaded from the
image servers, and a sample of galaxies was chosen. In the Coma
cluster, as many galaxies as possible in common with the sample of
Trager et al. (2008) were selected. In Virgo, galaxies were chosen
where possible to be in the sample of the ACS Virgo cluster survey
(C\^ot\'e et al. 2004) and the catalogue of Michard (2005), note
however that the majority of Michard's sample are too large for this
study. Galaxies were rejected from the sample if they fell too close
to an image boundary, to a neighbouring galaxy, or to a superimposed
star to allow definition of a clean circular aperture centred on the
galaxy. The final sample contained 14 galaxies, eight from Coma and
six from Virgo. In Table ~\ref{tab:properties} we present the basic
properties of this sample; galaxy types, redshifts and the Schlegel
extinctions are taken from the NASA Extragalactic Database (NED), and
the absolute V band magnitudes are taken from the Third Reference
Catalogue (RC3, de Vaucouleurs et al. 1991), assuming distance moduli
of 35.0 for Coma, and 31.1 for Virgo.

\begin{table}
\caption{\label{tab:properties}Basic properties of the sample galaxies.}
\begin{tabular}{ccccc}\hline
\bf{Galaxy}&\bf{Type}&\bf{Redshift}&\bf{$A_V$}&\bf{$M_V$}\\
&\bf{(NED)}&&\bf{(Schlegel)}&\bf{(RC3)}\\
\hline
IC3501&d:E1&0.0055&0.092&--17.37\\
NGC4318&E&0.0041&0.083&--17.88\\
NGC4515&S0-&0.0032&0.103&--18.60\\
NGC4551&E:&0.0039&0.128&--19.21\\
NGC4564&E6&0.0038&0.116&--20.10\\
NGC4867&E3&0.0162&0.034&--20.57\\
NGC4872&SB0&0.0241&0.030&--20.62\\
NGC4871&SAB0/a&0.0227&0.030&--20.89\\
NGC4873&SA0&0.0194&0.027&--20.93\\
NGC4473&E5&0.0075&0.094&--20.99\\
NGC4881&E0&0.0225&0.036&--21.47\\
NGC4839&cD&0.0246&0.032&--22.97\\
NGC4874&cD&0.0241&0.028&--23.35\\
NGC4889&cD/E4&0.0217&0.032&--23.54\\
\hline
\end{tabular}
{\it Notes to Table~\ref{tab:properties}:} Column 1 gives the galaxy
name, column 2 the type from NED. Column 3 gives the redshift, again
from NED, column 4 the extinction $A_V$ using the Schlegel (1998)
maps. Column 5 is a $V$ band absolute total magnitude for the galaxy
from RC3, using the distance modulus estimates given in the text.
The IC3501 $M_V$ is determined from interpolation of $g$ and $r$ flux 
densities. 
\end{table}

For each galaxy we measured aperture magnitudes in a series of
apertures using the {\sc iraf} task {\sc phot}. Sky levels were
determined in a number of regions of the image chosen to be free of
foreground stars. Galaxy centres were determined in each image using
the centroiding function within {\sc phot}.  For each galaxy we then
chose an optimum aperture, in the range 20 -- 40 arcsec diameter,
within which to measure the magnitude in all passbands. All apertures
are large enough that the different point spread
functions of SDSS and 2MASS data do not have a significant effect. The
optimum aperture chosen was in each case a compromise between the size
of the galaxy, the proximity of neighbours, and the additional noise
in the 2MASS magnitudes in larger apertures.  

 Magnitude zeropoints were taken directly from the fits 
headers of the 2MASS images, however for the SDSS data the zero 
point keyword ({\tt FLUX20}) in the fits header of the downloaded image files
(fpC files) has not been calculated
using the final calibration information\footnote{http://www.sdss.org/DR6/algorithms/fluxcal.html}.
Instead we recalibrated the fpC images using the zeropoint, extinction correction,
and airmass given in the tsField files for the relevant Run and Field number. 
This recalibration gives a substantial correction to the magnitudes 
calculated using the image headers, particularly in the $u$, $g$ and $z$ bands where 
the corrections are in the range 0.05 -- 0.1 magnitudes.

\subsection{Homogenising galaxy photometry}

In order to compare the models to our galaxy observations, the
measured magnitudes must be homogenised to the same photometric
system, for which we choose the AB magnitude system. Our optical
photometry is carried out in the SDSS DR6 system, however there are
sky position dependent differences between DR6 and DR7 zero points
(see Padmanabhan et al. 2008). We transform our photometry to the DR7
photometric system by adding the difference between zero points
calculated for each galaxy from the difference between DR6 and DR7
catalogue magnitudes. The SDSS DR7 photometry is brought onto the AB
magnitude system by subtracting 0.04 from the $u$ and adding 0.02 to
the $z$ band magnitudes as recommended on the SDSS web
pages\footnote{http://www.sdss.org/dr6/algorithms/fluxcal.html\#sdss2ab}. We
note that Eisenstein et al. (2006), in their work on hot white dwarfs,
recommend a slightly different AB transformation vector, the main
difference of which is to add approximately 0.015 to the $i$ band
magnitudes. This vector might improve marginally many of our fits,
however as the SDSS project appears not to have adopted the Eisenstein
et al. transformation vector we do not feel justified in doing so.

The Infrared 2MASS magnitudes make use of the Vega magnitude system;
these are brought on to the AB system by adopting the conversions from
Cohen, Wheaton \&\ Megeath (2003), shown in table \ref{vegajhk}. As a
check we also convolved the $\alpha$ Lyr\ae\ (Vega) spectral energy
distribution of Kurucz (1993) with the 2MASS $JHK_s$ response curves
and converted the results to AB magnitudes, in effect calculating the
magnitudes of Vega in the AB system. After taking into account the AB
magnitude of Vega (0.03), our values are in agreement with those of
Cohen, Wheaton \&\ Megeath (2003).

\begin{table}
\caption{Conversions between 2MASS (Vega) and AB magnitude systems for
  the $JHK$ bands, taken from Cohen, Wheaton \& Megeath (2003). To apply
  the corrections, $m_{AB} = m_{\rm Vega} + c$, where $c$ is the
  relevant correction from the table, shown to 3 decimal places. The
  uncertainties for each of the corrections are added in quadrature to
  the photometric errors in the $J$, $H$ \& $K_s$ bands.}
\label{vegajhk}
\begin{tabular}{l|ccc}
\hline
{\bf Band} & $J$ & $H$ & $K_s$ \\
\hline
{\bf Correction} & 0.894 & 1.374 & 1.840 \\
{\bf Uncertainty} & 0.022 & 0.022 & 0.022 \\
\hline
\end{tabular}
\end{table}

\subsection{Photometric errors and corrections}

For each magnitude we calculated an error, allowing for three sources
of uncertainty. The first was the error in the measurement process,
which in turn results from errors in the determination
of the sky level. The measurement error was
estimated from the SDSS and 2MASS images by repeated measurements 
of the magnitudes from the same image.  
In the 2MASS images this is a significant, albeit never dominant,
source of error. The second source
of error comes from noise in the image. This is calculated from the
image gains given in the fits headers, and the image statistics as
measured in the apertures in the case of SDSS passbands, and as given
in the image headers in the case of 2MASS. For the 2MASS data the
errors are corrected to allow for the resampling and smoothing applied
to the data before delivery, using the prescription given on the 2MASS
web
pages.\footnote{http://www.ipac.caltech.edu/2mass/releases/allsky/doc/sec6\_8a.html}

The third source of error is the error on the photometric zero point,
and this is taken from the web pages of
SDSS\footnote{http://www.sdss.org/dr5/algorithms/fluxcal.html} and
2MASS\footnote{http://www.ipac.caltech.edu/2mass/releases/allsky/doc/sec4\_8.html}
respectively. For the 2MASS data we consider also the uncertainty in the
conversion of 2MASS to AB magnitudes, as given in Table ~\ref{vegajhk}.
These error sources are all added in quadrature. The
zero point error always dominates for all SDSS passbands, whereas the
image noise error dominates in most cases for 2MASS bands.

Magnitudes are not K-corrected, as the fitting procedure we use
integrates the redshifted model flux over the rest wavelength filter
passbands, however they are corrected for Galactic extinction, using
the $A_V$ estimates of Schlegel et al.  (1998), as tabulated in NED,
and converted to SDSS passbands using the prescription of Kim and Lee
(2007), and to $JHK$ using the Schlegel et al. prescription which is in
turn derived from the functional form of Clayton \& Cardelli
(1988). The final magnitudes and errors used for the fitting are
presented in Table 3.

\begin{figure*}
\includegraphics[height=24cm]{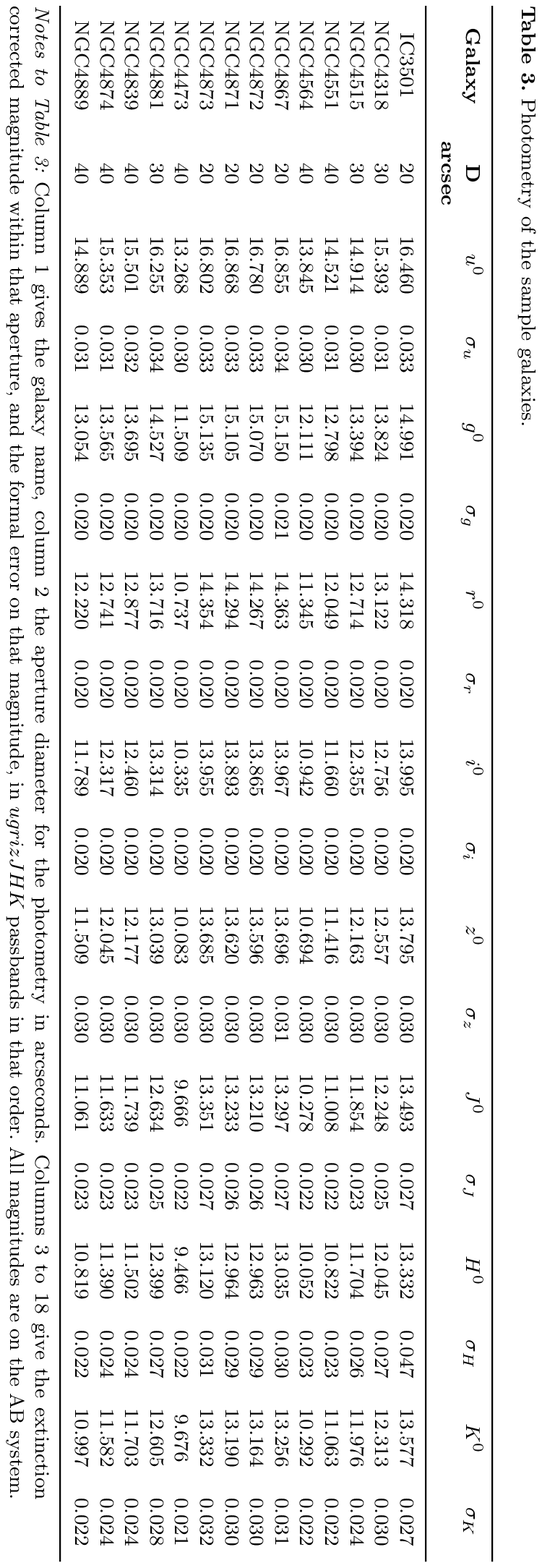}

\end{figure*}

\section{Stellar Population Models}

We have chosen to test seven popular sets of stellar population models
against the data presented in Section~\ref{sec:data}. All models
produce a series of synthetic spectra at points on a grid in
age--metallicity space which are compared with the broad-band
photometry as described in Section~\ref{sec:fitting}. For convenience,
the basic parameters of each model are presented in Table 4, and we
describe here briefly the further relevant details, in particular the treatment of the TP-AGB
stars which are generally not included in the adopted stellar
evolutionary tracks. We refer the reader to the original papers
describing these models for a full description of each model.

\begin{figure*}
\includegraphics[height=24cm]{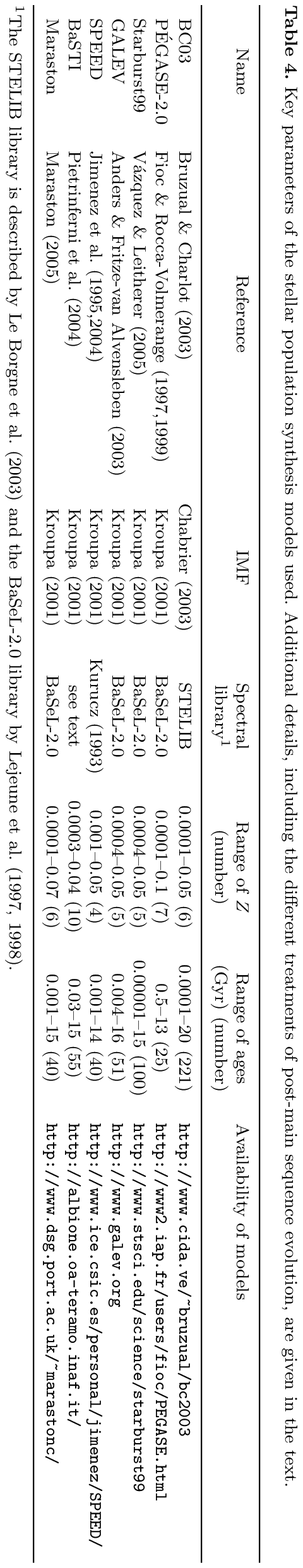}

\end{figure*}

\begin{enumerate}
\item Bruzual \& Charlot (2003) models (BC03) use the Padova 1994
  evolutionary tracks (Alongi et al.\ 1993; Bressan et al.\ 1993;
  Fagotto et al.\ 1994a,b; Girardi et al.\ 1996) with the 
  TP-AGB and Post Asymptotic Giant
  Branch (PAGB) phases treated according to Vassiliadis \& Wood (1993,
  1994). We use the STELIB spectral library in our modelling, although
  our results do not change if we use the BaSeL-2.0 library. Neither
  stellar library contains carbon stars or upper TP-AGB stars and the
  spectra for these stars are constructed from observational libraries
  as described by BC03.  

\item P\'{E}GASE models (Fioc \& Rocca-Volmerange 1997, 1999) also use
  stellar tracks from the Padova group, and pseudo-tracks for the
  TP-AGB phase as proposed by Groenewegen \& de Jong (1993).  

\item Starburst99 (V\'azquez \& Leitherer 2005) is of particular
  interest in the study of young stellar populations, although it is
  designed to model populations of any age. It uses the Padova 1994
  stellar tracks and incorporates TP-AGB stars in a similar manner to
  BC03, as described by V\'azquez \& Leitherer (2005).  Spectra were
  computed using the {\tt PAULDRACH/HILLIER} option, which uses the
  BaSeL-2.0 atmospheres for all except the hottest stars. Starburst99 
  models using Geneva high mass-loss isochrones were also computed,
  but provided a very poor fit. 

\item GALEV (Anders \& Fritze-van Alvensleben 2003) is an evolutionary
  synthesis code written to study the spectral and chemical evolution
  of galaxies, underpinned by a set of SSP models (Schulz et
  al.\ 2002; Anders et al.\ 2004) using Padova isochrones and BaSeL-2.0 atmospheres.

\item The SPEED models are described by Jimenez et al.\ (1995, 2004)
  and adopt a stellar evolution code based upon Eggleton (1971, 1972)
  but with improvements, particularly in the modelling of the HB and
  AGB. Mass loss at the RGB tip is modelled using a distribution of
  values of the Reimers (1975) mass loss parameter $\eta$, and this
  distribution is calibrated against the observed HB morphologies of
  Galactic star clusters (Jimenez et al.\ 1996). The AGB is modelled
  as described by Jorgensen (1991).  Spectra are largely taken from
  Kurucz (1993), but the authors compute their own LTE stellar
  atmosphere models for stars cooler than 4000\,K.

\item Maraston (2005) presents a set of models determined from an
  evolutionary synthesis code (Maraston 1998) based upon the tracks of
  Cassisi et al.\ (1997a,b; 2000) for main sequence stars, and upon
  the Fuel Consumption Theorem (Renzini \& Buzzoni 1986) for post-MS
  evolution, calibrated against clusters in the Magellanic Clouds.

The BaSeL2.0 library of stellar spectra is used, with TP-AGB stars
taken from Lan\c{c}on \& Mouhcine (2002). Maraston's models are
available with two horizontal branch morphologies (RHB and BHB) and we
primarily consider the RHB models which are more appropriate for
metal-rich populations. However, the existence of a blue HB morphology
in a small number of metal-rich Galactic globular clusters (Busso et
al.\ 2007) leads us to consider the BHB models as well.
 
Maraston et al.\ (2008) discuss a set of related models which use the
Pickles (1998) stellar library instead of BaSeL-2.0, but these were
not available to us at the time of writing.

\item The BaSTI (Bag of Stellar Tracks and Isochrones) models are
  based on the work of Pietrinferni et al.\ (2004), extended to cover
  TP-AGB stars by Cordier et al.\ (2007).  Model spectra are
  constructed from the BaSTI isochrones as described by Percival et
  al. (2008), using the `low resolution' version of the model
  spectra. These are based upon the model atmospheres of Castelli \&
  Kurucz (2003), supplemented by models from the BaSeL-3.2 library
  (Westera et al. 2002) for cool stars (T $<$ 3500\,K), and by the
  empirical spectra of Lan\c{c}on \& Mouhcine (2002) for AGB carbon
  stars.

Because these models are `in-house', we are able to test different
prescriptions for $\alpha$-enhancement and HB morphology. In addition
to the scaled-solar models (Pietrinferni et al.\ 2004) we analyse also
the [$\alpha$/Fe] = +0.4 models and isochrones of Pietrinferni et
al.\ (2006). We also analyse two different values (+0.2 and +0.4) of
the Reimers (1975) mass-loss parameter, $\eta$. Discussion of the
effects of varying [$\alpha$/Fe] and $\eta$ is presented in Section
\ref{sect:alphaenhancement}

\end{enumerate}

\section{Integration and Fitting Procedures}
\label{sec:fitting}

\subsection{Calculating model fluxes}

In order to compare our observations with the results of SSP
modelling, we convert from our observed AB magnitudes and magnitude
errors to fluxes. To extract comparable fluxes, $S_n$, from the SSP
SEDs of each species there are a number of steps. First, we multiply
the wavelength values of the given SSP model by a factor of $(1+z)$ to
account for the redshift of the galaxy in question. Second we rebin
linearly the transmission function of each filter, $T_{\lambda,n}$,
onto the wavelength scale of the models.

For each SSP SED $f_\lambda$, filter $n$, and filter transmission
function $T_{\lambda, n}$, we then integrate the total flux in the passband
using trapezoidal summation:


\begin{equation}
S_n = \frac{\displaystyle\int f_\lambda \lambda T_{\lambda,n} d\lambda}
{\displaystyle\int \lambda T_{\lambda,n} d\lambda}
\label{modelflux}
\end{equation}

Next, we convert the observed AB magnitudes to 
$f_\lambda$ using:

\begin{equation}
f_{\lambda,n} = 10^{(-m_{AB}/2.5)} \times f_0 \times {c / \lambda_{eff,n}^2}
\label{eq:abconv}
\end{equation}

\noindent where $f_0$ is the AB magnitude zero point of 3631 Jy, and
$\lambda_{eff,n}$ is the effective, or pivot, wavelength of the passband, defined
by:

\begin{equation}
\lambda_{eff}^2 = {\int{T_{\lambda} \lambda d\lambda} \over \int{{T_{\lambda} \over \lambda} d\lambda}}
\end{equation}

\noindent We are now in a position to compare our observations with
the model SSP fluxes.

\subsection{$\chi^2$ fitting}

In order to quantify the extent to which the galaxies in our sample
are described by the array of models mentioned in section 3, we
employed a simple reduced $\chi^2$ test. For a given galaxy with
observed fluxes $f_i$ with each flux having an associated error
$\sigma_i$ in $n$ bands, the extent to which they are well described
by a model with fluxes $S_i$ is approximated by the reduced $\chi^2$,
calculated as follows;

\begin{equation}
\chi^2_{\nu} = \frac{1}{(n-\kappa)} ~\sum_{i=0}^{n} \frac{(NS_i - f_i)^{2}}{\sigma_i^2},
\label{chi2}
\end{equation}

\noindent where $N$ represents the normalisation of the model
magnitudes and $\kappa$ represents the number of constraints
associated with the models. The normalisation factor $N$ is calculated
by analytically solving the equation ${\rm d}\chi^2_{\nu}/{\rm d}N=0$. The
value of $\kappa$ used for these calculations is 3, due to the
treatment of a galaxy model's age, metallicity, and flux density
normalisation as free parameters. Model SSPs that approximate our
galaxy observations well will then have values of reduced $\chi^2$
approaching unity, while for those models which do not approximate our
observations well the values will be higher, depending on the
`goodness of fit'.

In order to discriminate between the large numbers of models for a
given galaxy, the best fit value of $\chi^2_{\nu}$ is calculated
for every SSP in each model species (e.g. BC03, M05, etc.). In this
way we build up a database of best-fit $\chi^2_{\nu}$ for all values
of SSP age and metallicity for each galaxy in our sample. Making use
of this database, we can then compare the results of our $\chi^2$
fitting to those of other studies (e.g. Trager et al., 2008). The
properties of the best fit models to each of the galaxies in our
sample are shown in tables 5 and 6.

\begin{figure*}
\includegraphics[height=24cm]{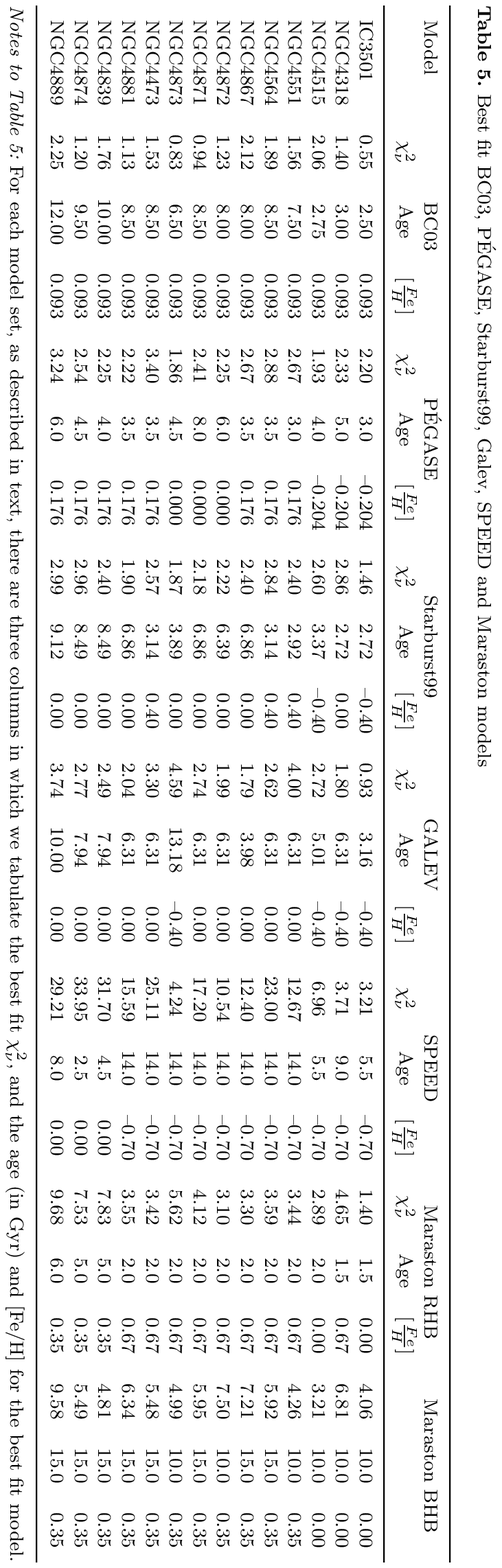}

\end{figure*}

\section{Analysis}

\begin{table*}
\caption{BaSTI models for different $\alpha$-enhancement and mass loss parameters}
\label{tab:basticomparison}
{\small
\begin{tabular}{l|ccc|ccc|ccc|ccc}
\hline
 & \multicolumn{12}{c}{BaSTI} \\
 & \multicolumn{3}{c}{$[{\alpha}/Fe]$ = 0.0, $\eta$ = 0.2} & \multicolumn{3}{c}{$[{\alpha}/Fe]$ = 0.0, $\eta$ = 0.4} & \multicolumn{3}{c}{$[{\alpha}/Fe]$ = 0.4, $\eta$ = 0.2} & \multicolumn{3}{c}{$[{\alpha}/Fe]$ = 0.4, $\eta$ = 0.4} \\
 & $\chi^2_{\nu}$ & Age & $[{Fe \over H}]$ & $\chi^2_{\nu}$ & Age & $[{Fe \over H}]$ & $\chi^2_{\nu}$ & Age & $[{Fe \over H}]$& $\chi^2_{\nu}$ & Age &  $[{Fe \over H}]$ \\
\hline
IC3051&1.61&4.50&--0.25&1.47&4.50&--0.25&1.04&8.50&--0.60&0.95&14.00&--0.60 \\
NGC4318&1.65&8.00&--0.25&1.58&8.00&--0.25&1.56&13.50&--0.60&2.27&5.50&--0.29 \\
NGC4515&2.66&6.00&--0.25&2.56&6.00&--0.25&2.67&10.00&--0.60&2.87&10.50&--0.60 \\
NGC4551&2.62&2.75&0.40&2.50&2.75&0.40&2.76&7.50&--0.09&2.69&8.00&--0.09 \\
NGC4564&2.78&3.00&0.40&2.71&3.00&0.40&1.31&6.50&0.05&1.40&10.00&--0.09 \\
NGC4867&2.91&3.00&0.40&2.82&3.00&0.40&0.54&6.00&0.05&0.57&6.00&0.05 \\
NGC4872&2.43&3.00&0.40&2.37&3.00&0.40&0.73&8.50&--0.09&0.74&9.00&--0.09 \\
NGC4871&2.09&3.00&0.40&2.07&3.00&0.40&0.89&10.00&--0.09&0.86&10.00&--0.09 \\
NGC4873&2.05&3.00&0.26&2.05&3.00&0.26&0.91&11.00&--0.29&0.99&11.00&--0.29 \\
NGC4473&2.37&3.00&0.40&2.33&3.00&0.40&2.36&6.50&0.05&2.39&10.00&--0.09 \\
NGC4881&2.60&3.00&0.40&2.54&3.00&0.40&0.77&6.50&0.05&0.78&10.00&--0.09 \\
NGC4839&3.90&6.00&0.40&2.70&5.00&0.40&1.17&8.50&0.05&1.18&8.50&0.05 \\
NGC4874&3.38&4.50&0.40&2.20&5.00&0.40&1.28&11.50&--0.09&1.31&12.00&--0.09 \\
NGC4889&4.46&6.50&0.40&4.37&6.50&0.40&1.34&10.00&0.05&1.35&10.50&0.05 \\
\hline
\end{tabular}}

{\it Notes to Table ~\ref{tab:basticomparison}:} {Columns 2 -- 13} - Best fitting BaSTI models for two values of $[{\alpha}/Fe]$, and
for two values of the Reimers (1975) mass loss parameter $\eta$. For each combination of parameters, there are
three columns in which we tabulate the best fit $\chi^2_{\nu}$, and the age (in Gyr) and [Fe/H] for the best fit model.
\end{table*}

Remembering always that we are observing potentially composite
populations, and comparing them with SSP models, we examine in turn
the model sets, and how well they fit to the data.

\begin{figure*}
  \begin{minipage}[c]{8cm}
    \vspace{0pt}
    \includegraphics[width=8.0cm]{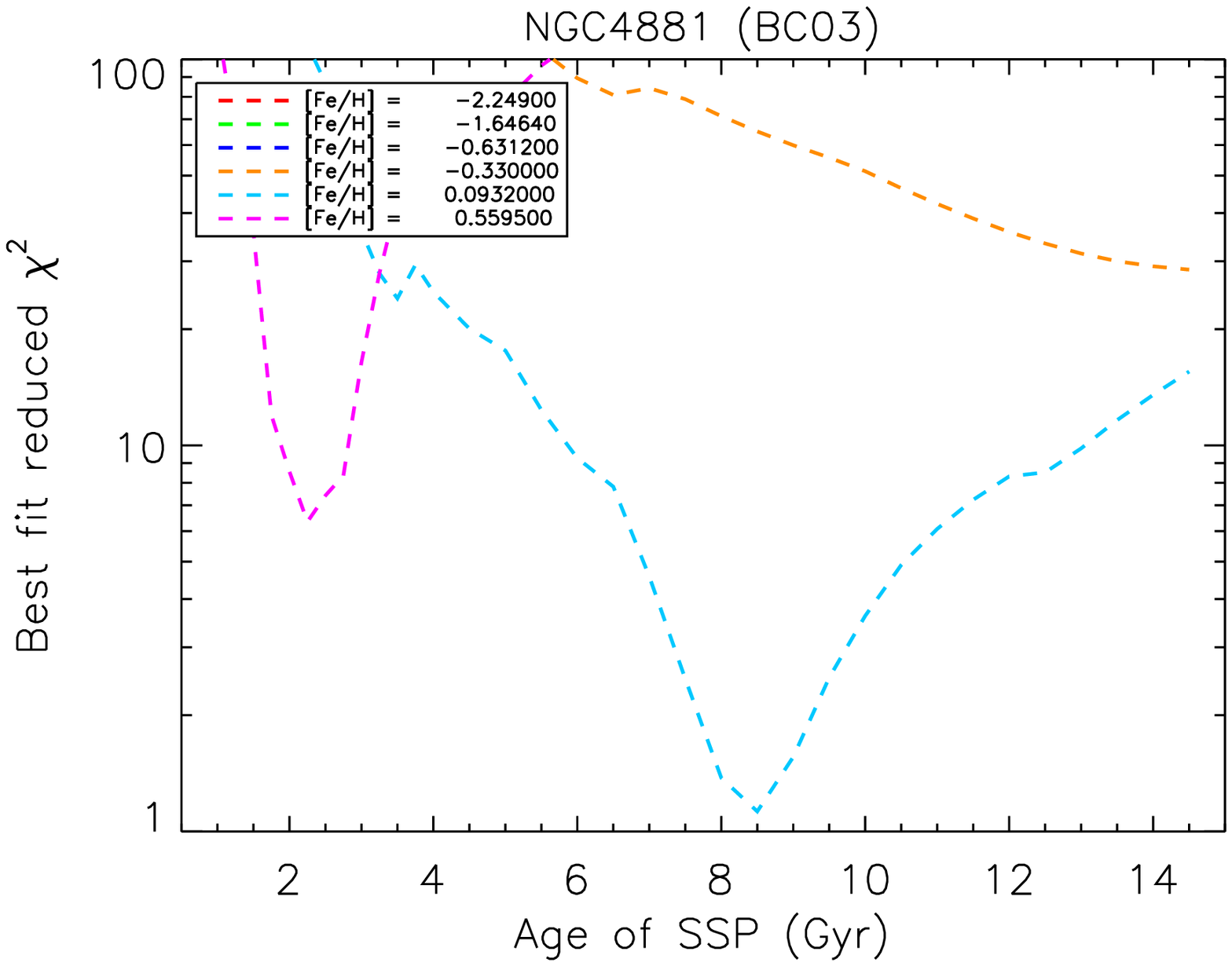}
  \end{minipage}
  \hfill
  \begin{minipage}[c]{9cm}
    \vspace{0pt}
    \includegraphics[width=9.0cm]{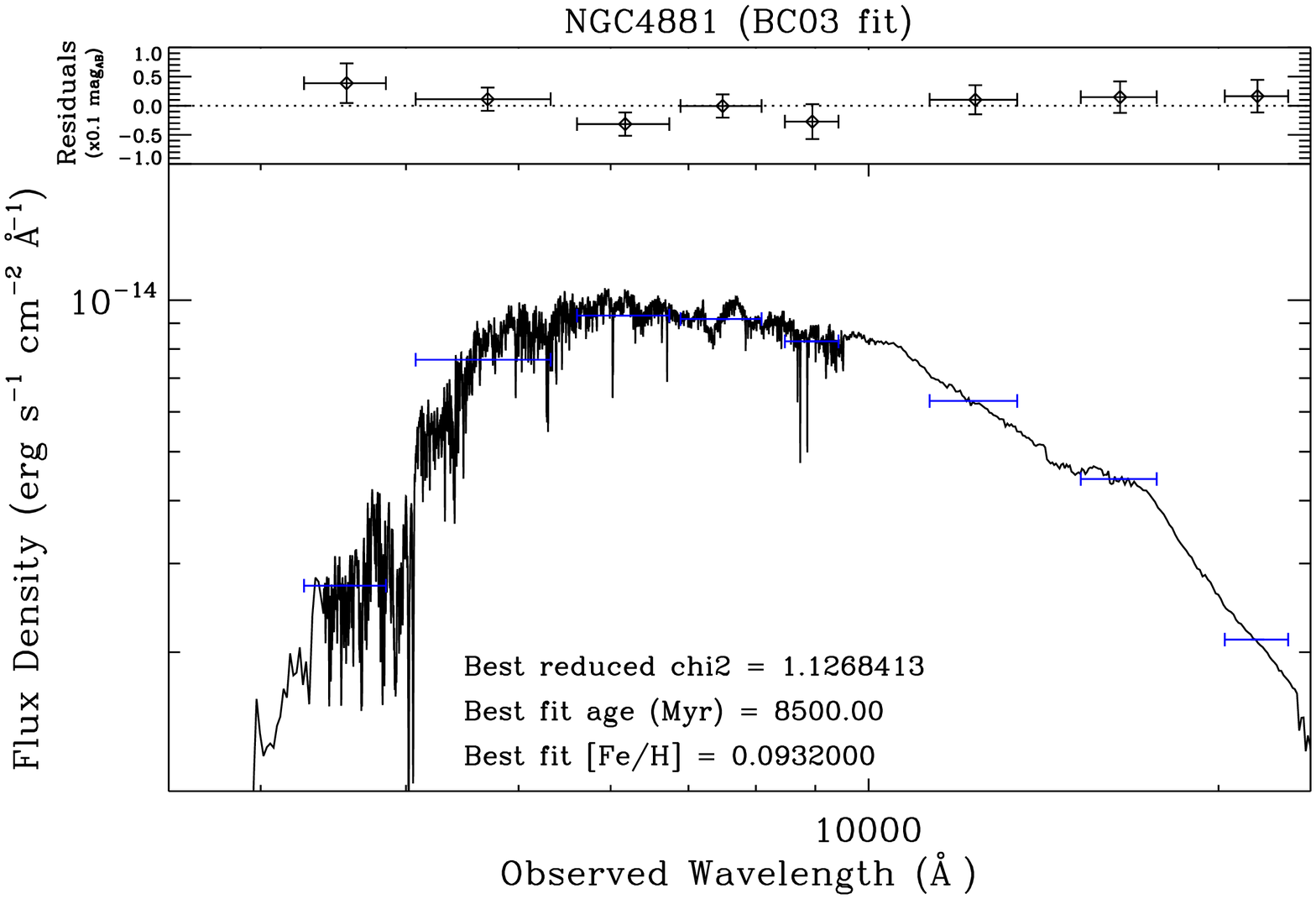}
  \end{minipage}
\caption{Left panel: $\chi^2_{\nu}$ against age for BC03 models of different
  metallicities as fit to the photometry of NGC4881. Lower right
  panel: the best fit model spectrum ([Fe/H] = 0.093; Age = 8.5~Gyr)
  with the SDSS and 2MASS passbands overplotted (blue horizontal error
  bars).  Upper right panel: shows the residuals from the fit (data --
  model) in units of 0.1 magnitude}
\label{fig:NGC4881_bc03fit}
\end{figure*}

\begin{enumerate}
\item BC03 models. The BC03 models often provide the lowest values of
  $\chi^2_{\nu}$ and thus objectively the best fit for the low luminosity
  Virgo galaxies in particular. In all cases this best fit comes from
  the [Fe/H] = 0.093 model, at an age between 2.5 and 12~Gyr, with the
  massive Coma galaxies in general showing the oldest ages.  The
  residuals around the fit show a systematic pattern, in particular
   ($r-i$) is predicted too blue by 0.01 - 0.03 magnitudes for the
  lower luminosity Virgo galaxies, and 0.03 - 0.05 magnitudes for the
  more massive Coma galaxies. BC03 models provide too coarse a
  metallicity grid to investigate age-metallicity degeneracy, and in
  particular their one supersolar metallicity is [Fe/H] = 0.559, which
  is apparently too metal rich to fit any of our sample.

  In Figure ~\ref{fig:NGC4881_bc03fit} we show the dependence of
  $\chi^2_{\nu}$ on age, the best-fit model spectrum, and the residuals
  between data and best fit model, all for NGC4881, one of the best
  fitting galaxies. Similar plots for all galaxies and all model sets
  are available in the online data accompanying this paper.
\item P\'{E}GASE models. These models provide higher values of
  $\chi^2_{\nu}$ than BC03, usually because they underpredict the flux in
  the $i$ band.  P\'{E}GASE models predict ($r-i$) to be too blue, and ($i-z$)
  to be too red, both by as much
  as 0.1 magnitude. The solar metallicity models show a minimum at
  generally slightly younger ages than BC03, but there is quite often
  a better fit from an interpolated model at [Fe/H] = --0.204 or
  +0.176. The $\chi^2_{\nu}$ curves for these models show clearly the
  age-metallicity degeneracy, where the supersolar model gives a fit a
  factor 1.5 -- 2 younger, and the subsolar one a factor of
  $\sim$2 older.
\item Starburst99 models. Using Padova isochrones, these models give
  very similar results to P\'{E}GASE, in that they predict the same
  ($r-i$) and ($i-z$) discrepancy, they show very similar ages, and the same
  age-metallicity degeneracy. They have less metallicity resolution
  than P\'{E}GASE, and sometimes this results in a higher best value
  of $\chi^2_{\nu}$. Starburst99 models using Geneva
  isochrones fit substantially less well in all cases, and we do not
  consider these models further.
\item GALEV models. GALEV in general does not give quite as good a fit
  as P\'{E}GASE, but the best fits usually occur at a similar age and
  metallicity. GALEV ages are similar to P\'{E}GASE ages, and in
  general older than Starburst99 ages. These models again predict both 
  ($r - i$) to be too blue and ($i-z$) to be too red, by 0.05 -- 0.1 magnitude.
\item SPEED models. The best fit SPEED models are almost always old
  (often at their limit of 14 Gyr), and metal-poor (Z=0.001 or 0.004).
  They also overpredict the $u$ band flux by 0.1 mag or more, and as
  with other model sets predict ($r$ - $i$) to be too blue by 0.05 -
  0.1 mag. The $\chi^2_{\nu}$ values are consequently high compared with
  other models.
\item Maraston (2005) models. The RHB models provide a less good fit
  to the broad band data than BC03, P\'{E}GASE, Starburst99 or BaSTI,
  with $\chi^2_{\nu} < 3$ for only two of the fourteen galaxies. The general
  pattern of the residuals is that ($r-i$) is predicted too blue, by 0.15
  magnitude. For the Coma galaxies ($J-K$) is also too blue, by around
  0.05 magnitude. The best fit ages are also uncomfortably young, in
  the range 1.5 -- 3 Gyr even for massive cluster ellipticals. Such ages
  are in conflict with those derived from line indices (Trager et
  al. 2008). The BHB models are only appropriate for ages $>$ 10~Gyr 
  and would generally be expected to fit low metallicity populations, 
  so it is somewhat surprising that the [Fe/H] = 0.35 BHB model 
  provides a better fit to the data than any 
  of the RHB models, for the massive Coma galaxies in particular.
\item BaSTI models. The standard scaled-solar abundance models give
  slightly higher values of $\chi^2_{\nu}$ than BC03, and similar to
  P\'{E}GASE or Starburst99. The residuals show an overprediction of
  the $r$ band flux, of 0.03 - 0.08 magnitudes. As with P\'{E}GASE, BaSTI gives younger
  ages than BC03, largely because it has two supersolar metallicities
  ([Fe/H] = +0.26 and +0.40). In section ~\ref{sect:alphaenhancement}
  we discuss the effect of making different assumptions about
  [$\alpha$/Fe] and the Reimers (1975) mass loss parameter $\eta$,
  noting that we can only alter these parameters with the BaSTI
  models.
\end{enumerate}

\begin{figure*}
\includegraphics[width=12cm]{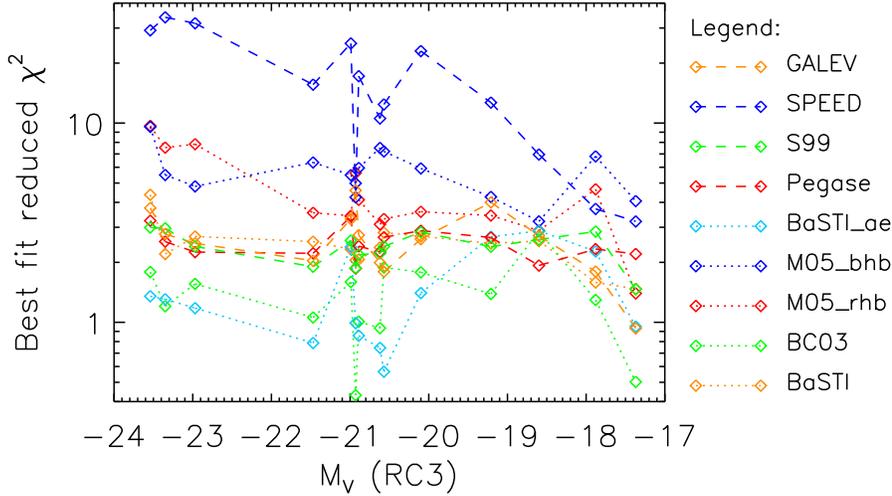}
\caption{Best fit $\chi^2_{\nu}$ against absolute magnitude for each model set.}
\label{fig:vabs_bestfit}
\end{figure*}

In Figure ~\ref{fig:vabs_bestfit} we show $\chi^2_{\nu}$ at the best fit age
and metallicity against absolute magnitude for our sample. There are
some trends in some model sets, for instance the $\alpha$-enhanced
models perform best for more luminous (and massive) galaxies, whereas
BC03, P\'{E}GASE, Starburst99 and BaSTI scaled-solar models perform
better at lower luminosity.

\subsection{Age-metallicity degeneracy}

BaSTI and P\'{E}GASE models have sufficient metallicity resolution
that for most galaxies, there are two, three or even four
metallicities for which an age solution can be found with
$\chi^2_{\nu}$ within 1.0 of the lowest. Table 7
presents these solutions.  They have very different ages, and show the
well known age-metallicity degeneracy in the sense that the highest
metallicities give the youngest ages.

\begin{figure*}
\includegraphics[height=24cm]{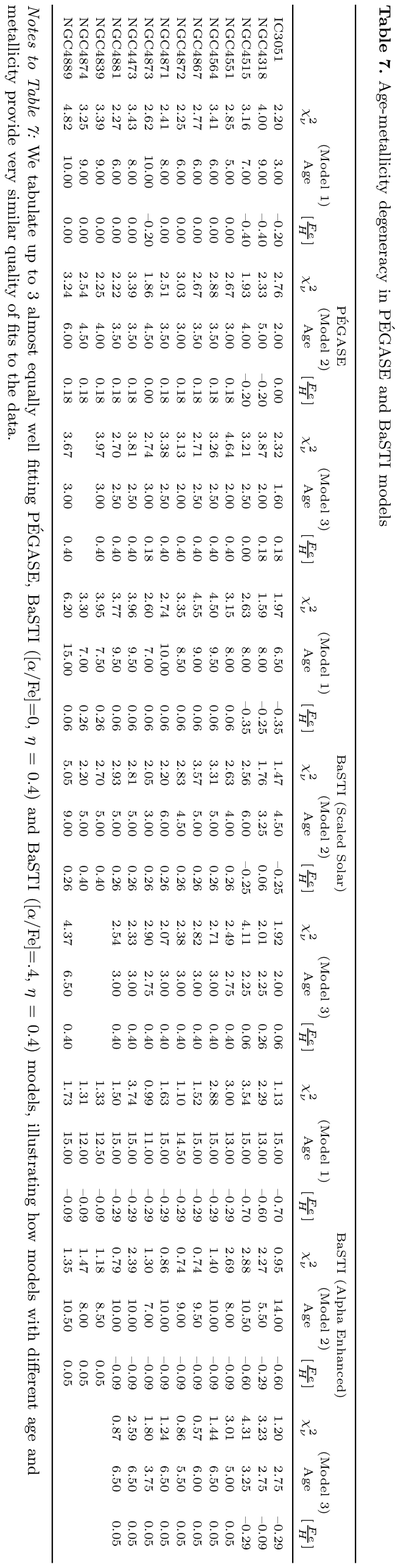}

\end{figure*}

\begin{figure*}
  \begin{minipage}[c]{8.5cm}
    \vspace{0pt}
    \includegraphics[width=8.5cm]{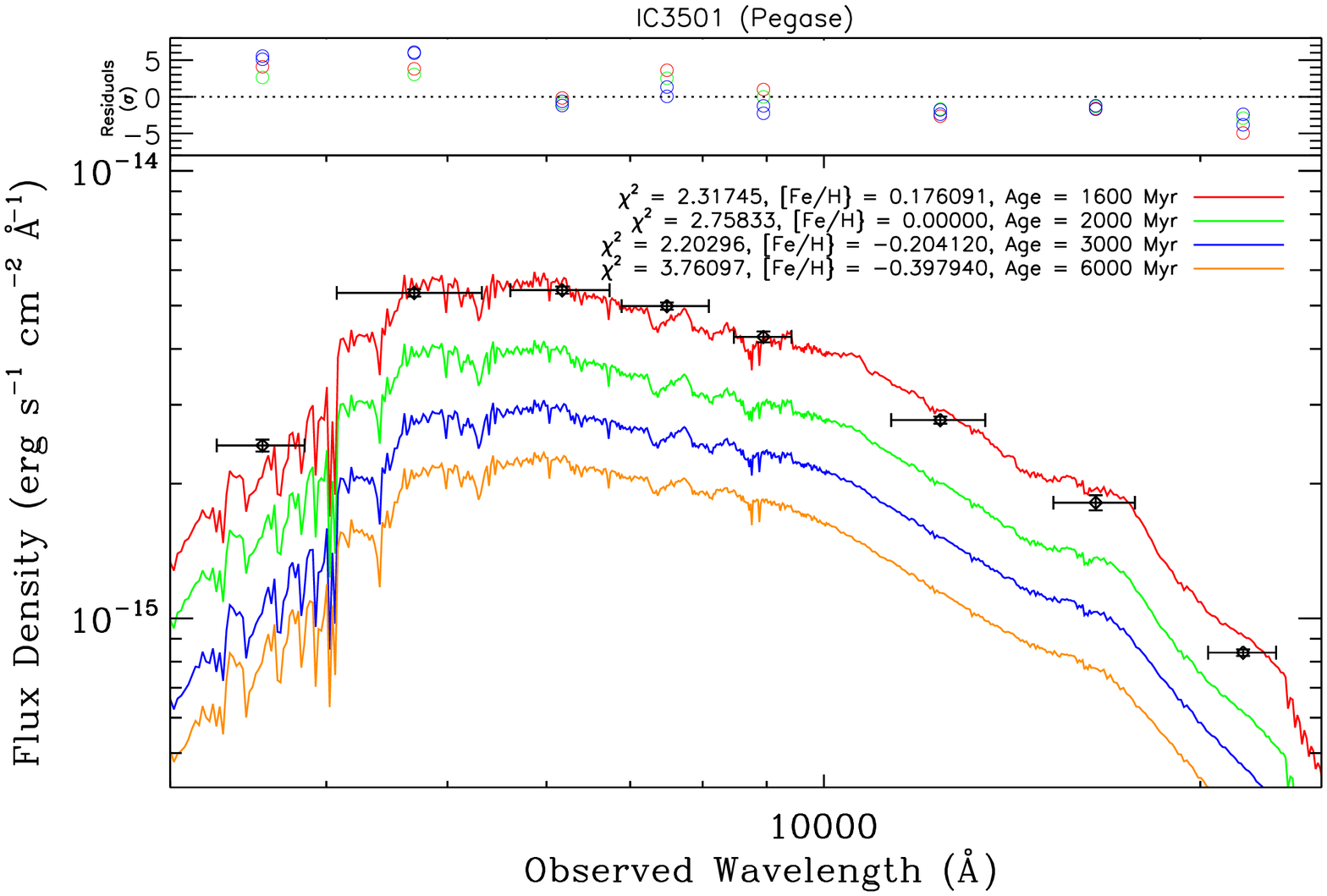}
  \end{minipage}
  \hfill
  \begin{minipage}[c]{8.5cm}
    \vspace{0pt}
    \includegraphics[width=8.5cm]{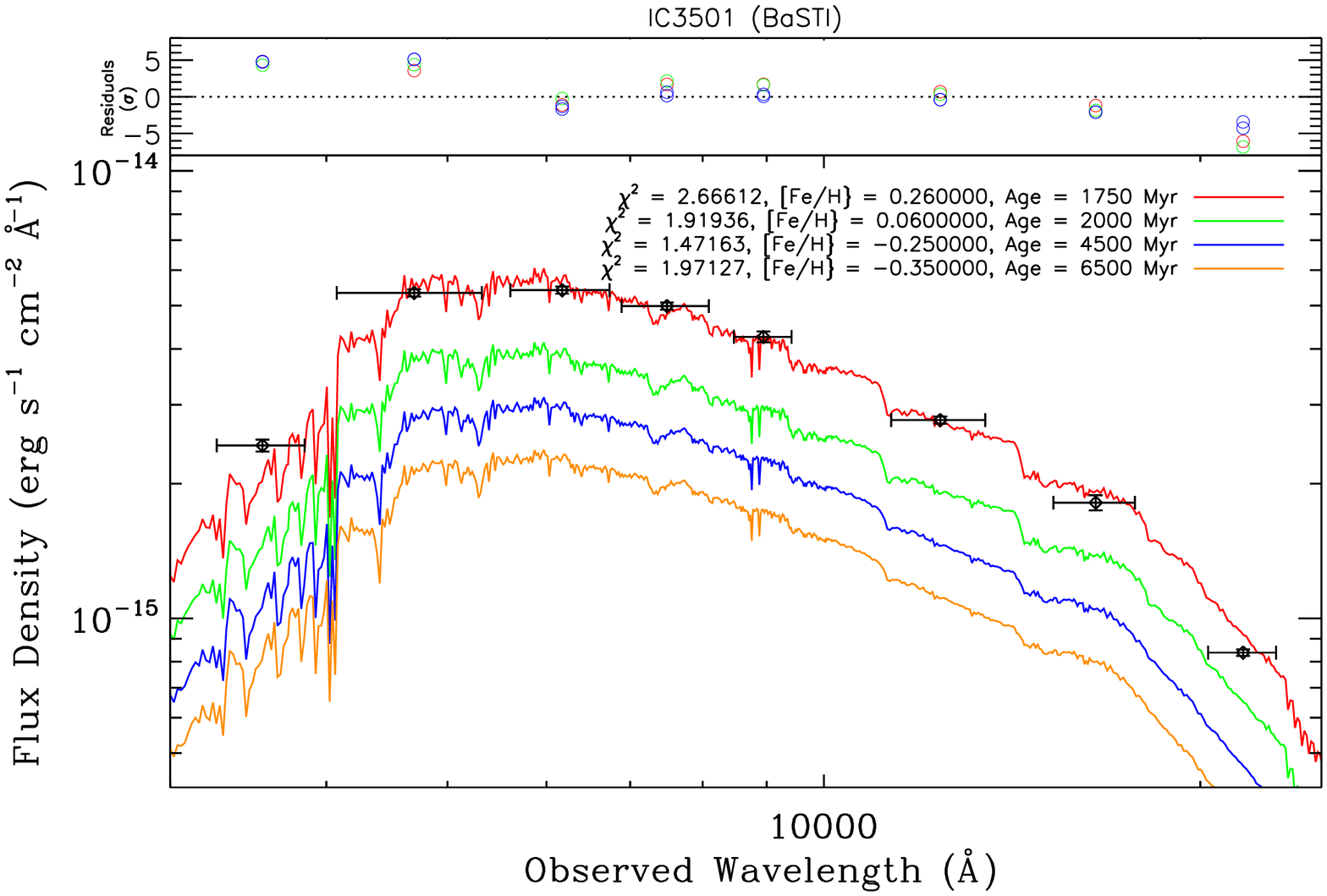}
  \end{minipage}
  \caption{Illustration of the age-metallicity degeneracy, we show the
  best age fits at four metallicities for IC3501, using P\'{E}GASE
  models (left panel) and BaSTI scaled-solar models (right panel). In each case the three
  older and more metal poor models are offset vertically, for clarity.}
\label{fig:degeneracy}
\end{figure*}

In Figure ~\ref{fig:degeneracy} we illustrate the close correspondence
of the models of different age and metallicity for IC3501.  For each
model set, three models plotted in different colours, and shifted
vertically to separate the traces. The values of [Fe/H] = --0.204 for P\'{EGASE} and [Fe/H] = --0.25
for BaSTI provide the best fits.  $\chi^2_{\nu}$ in each case is within 1 of
the best fit, so the outer models, covering a metallicity range of
$\simeq$ 0.4 in [Fe/H], and an age range of a factor two, cannot be
ruled out.

Independent of model set, we find the age and metallicity changes to be
linked by the following relation:

$${{\delta{\log{Age}} \over {\delta{[Fe/H]}}} \simeq 1.2 \pm 0.25}$$

Without other information to constrain metallicity or age, it appears
that using broad band photometry from $u$ to $K$, we can measure these
to $\Delta{[Fe/H]} \simeq 0.18$ and $\Delta{\log{Age}} \simeq 0.25$.
Formal errors on the fitted age may be less than this for those model
sets with a coarser metallicity grid, but these errors mean very
little. Metallicities and ages can be constrained better by using line
indices, with e.g. Trager et al. (2008) quoting errors of 0.02 -- 0.15
dex on metallicity values and 0.04 - 0.20 dex on log(age) values
derived using line indices alone.

The substantial errors on ages are disappointing, given the earlier
claims that pairs of, for example, optical - near-infrared colours can
break the age-metallicity degeneracy.  Analysis of the colour-colour
grids reveals why substantial errors should be expected, particularly
for old stellar populations.  In the BaSTI simple stellar population
grids of $B-K$ vs. $J-K$ used by James et al. (2006), the $B-K$
colour, which is the principal age indicator, changes more between 3
and 5~Gyr than it does between 5 and 14~Gyr, and over the range
10-14~Gyr, the change in $B-K$ is typically only 0.02 -- 0.03 mag.  In
addition, for high metallicity and old populations, increasing age
leads to redder $J-K$ colours, so while the grid is not degenerate,
the age and metallicity vectors are far from orthogonal.  The same
trends are found for the $V-I$ vs. $V-K$ grids of Puzia et al. (2002),
which show a larger change in colours from 1 - 2~Gyr than from 5 --
15~Gyr.  Thus, while such techniques can be very sensitive to the
presence of even a small mass fraction of young or intermediate-age
stars, the age discrimination for old populations is poor given
typical photometric errors.  Passbands blueward of our range may help
(Kaviraj et al. 2006), as the models presented in Table
7 deviate from each other blueward of $u$, but
uncertainties due to modelling of the Horizontal Branch become more
serious.

This analysis emphasises, on the other hand, the importance of 
having models with sufficient age and metallicity resolution to identify
the uncertainties in the derived parameters, and to investigate whether,
for instance, the quality of the fits distinguishes between the fundamental
parameters of the model sets, or whether the data demand more than an
SSP to fit.

\subsection{ Abundance ratios and the morphology of the Horizontal Branch}
\label{sect:alphaenhancement}

\begin{figure}
  \begin{minipage}[c]{8.5cm}
    \vspace{0pt}
    \includegraphics[width=8.5cm]{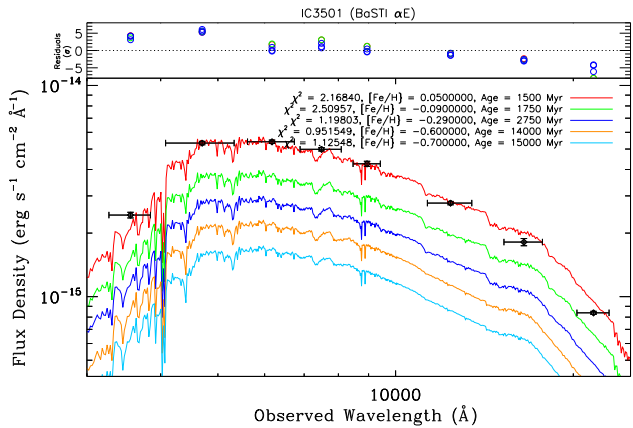}
  \end{minipage}
\caption{Age-metallicity degeneracy in the $\alpha$-enhanced models,
  we show the best age fits at five metallicities to IC3501. This is an
  extreme case, with an allowed age range of a factor 10, due to the
  transition in HB morphology at ages $>$ 8 Gyr in the high $\eta$,
  high [$\alpha$/Fe], low metallicity models. Again the older and more
  metal poor models are shifted vertically for clarity.}
\label{fig:degeneracy_ae}
\end{figure}

The BaSTI [$\alpha$/Fe] = +0.4 models are the only $\alpha$-enhanced
models that we investigate. For the Coma galaxies they provide better
fits than any scaled solar models, occasionally excepting BC03, but
for the lower-luminosity Virgo galaxies they are less successful. The
pattern of residuals shows that the overprediction in $r$ is less than
for the scaled-solar models, although they still predict ($r-i$) to be
too blue, by 0.02 - 0.05 magnitude.  However the main discrepancies
are in ($u-g$), which is predicted too blue by 0.05 -- 0.1
magnitude. The age-metallicity degeneracy still exists, with the three
most metal rich models ([Fe/H] = --0.29, --0.09 and +0.05) showing
minima with very similar values of $\chi^2_{\nu}$, but at ages which cover a
range of a factor three, as can be seen in the fits to NGC4881 in the
right panel of Figure ~\ref{fig:degeneracy_ae}. An extreme example can
be seen in the left panel of this Figure, which shows the fits for
IC3501.  The metal-poor ([Fe/H]= --0.60 and --0.70), [$\alpha$/Fe] =
0.4; $\eta$ = 0.4 models fit very well at old ages due to a sharp
transition from red to blue HB morphology at 8 -- 14 Gyr. For other
models this transition occurs at ages greater than the expected age of
the universe. Using our broad-band photometry alone these models are
almost indistinguishable from the $\sim$ 2 Gyr, [Fe/H] $\sim$ 0.29
models in this set (as Figure ~\ref{fig:degeneracy_ae} shows), and
from moderately metal-poor scaled-solar models. Similar effects are
seen in the fits to NGC4318 and NGC4515.

In general, the best fit ages for the $\alpha$-enhanced models are
greater than scaled-solar, with $\Delta({\rm log} Age) = 0.34 \pm
0.12$, a mean factor of two.

The Reimers (1975) mass-loss parameter $\eta$ has less effect, apart
from in the metal-poor $\alpha$-enhanced models for the lowest mass
galaxies discussed above. In other cases, when different metallicities
have very close minimum $\chi^2_{\nu}$, changing $\eta$ will occasionally
favour a different solution which, because of the age-metallicity
degeneracy, will then have different age. For the [$\alpha$/Fe] = +0.4
models the $\eta$ = 0.2 solution can be one age step (500 Myr) younger
at the same metallicity.

Maraston (2005) BHB models have very high values of $\eta$, between
0.45 and 1.0, and are designed to have a blue HB morphology even for
metal-rich (0.5 -- 2.0 solar) metallicities. These are available for
very old populations (10 -- 15 Gyr). The ages of the best fits are
constrained to be old, and the best fit metallicities are in the
range [Fe/H] = --0.33 to +0.35 corresponding well with values derived
from line indices. However the $\chi^2_{\nu}$ values are still high compared
with other model sets. These models all underpredict the $i$ band flux
by around 0.1 magnitude, and have mixed success at predicting ($u$ -
$g$), which in the more massive galaxies is predicted too red by 0.1
mag.

\section{Discussion}

\subsection{Effect of extinction estimates}

We have applied the extinction corrections of Schlegel et al. (1998)
but it is important to estimate the uncertainty in the derived
population parameters which results from errors in the extinctions. To
estimate the effect of uncertainties in the extinction estimates, for
a limited set of the stellar models, we have calculated the best fit
age, metallicity and $\chi^2_{\nu}$, assuming that $A_V$ is 0.1 magnitude
above and below the Schlegel et al. value. In some cases this would imply
a negative $A_V$, however we use these values only to estimate the effect
of errors in $A_V$ upon the derived parameters.

\begin{table*}
\caption{Effect of different extinction estimates on the derived parameters for BC03 models}
\label{tab:extinctiontest}
{\small
\begin{tabular}{l|ccc|ccc|ccc}
\hline
   & \multicolumn{3}{c}{$A_V$ Schlegel-0.1}& \multicolumn{3}{c}{Schlegel extinction} & \multicolumn{3}{c}{$A_V$ Schlegel+0.1} \\
 & $\chi^2_{\nu}$ & Age & $[{Fe \over H}]$ & $\chi^2_{\nu}$ & Age & $[{Fe \over H}]$ & $\chi^2_{\nu}$ & Age & $[{Fe \over H}]$ \\
\hline
IC3051&1.23&2.75&0.093&0.55&2.50&0.093&0.50&2.00&0.093 \\
NGC4318&0.71&4.50&0.093&1.40&3.00&0.093&1.29&2.50&0.093 \\
NGC4515&1.80&3.00&0.093&2.06&2.75&0.093&2.86&2.25&0.093 \\
NGC4551&1.74&9.00&0.093&1.56&7.50&0.093&1.39&5.50&0.093 \\
NGC4564&2.54&11.00&0.093&1.89&8.50&0.093&1.78&7.00&0.093 \\
NGC4867&2.30&2.75&0.560&2.12&8.00&0.093&1.88&6.00&0.093 \\
NGC4872&1.62&10.00&0.093&1.23&8.00&0.093&0.93&6.00&0.093 \\
NGC4871&1.20&11.00&0.093&0.94&8.50&0.093&1.01&7.00&0.093 \\
NGC4873&1.19&8.50&0.093&0.83&6.50&0.093&0.43&4.50&0.093 \\
NGC4473&1.87&11.00&0.093&1.53&8.50&0.093&1.59&7.00&0.093 \\
NGC4881&1.56&10.50&0.093&1.13&8.50&0.093&1.05&7.00&0.093 \\
NGC4839&2.55&14.50&0.093&1.76&10.00&0.093&1.56&8.00&0.093 \\
NGC4874&1.89&13.50&0.093&1.20&9.50&0.093&1.20&8.00&0.093 \\
NGC4889&3.55&3.25&0.560&2.25&12.00&0.093&1.78&9.00&0.093 \\
\hline
\end{tabular}}

{\it Notes to Table ~\ref{tab:extinctiontest}:} Effect of extinction
estimate upon the results for BC03 models. We present the derived
parameters under the assumptions of the Schlegel et al. extinctions
presented in Table 1 and adopted throughout this paper, and also for
$A_V$ values 0.10 magnitudes above and below these. Ages are in Gyr.
\end{table*}

In Table \ref{tab:extinctiontest} we present the results for the BC03
models. BCO3 has a coarse metallicity grid, and only in one case does
a change of extinction correction affect the best fit
metallicity. For the remaining cases, there is a clear and
unsurprising degeneracy between the extinction correction and
age. Assuming an additional 0.1 magnitudes above the Schlegel et al.
value leads to a derived age a factor $\sim$1.3 lower than the nominal
value. However the Galactic foreground extinctions are not thought to
be this uncertain, and Burstein \& Heiles (1982) extinctions for our
galaxies differ by less than 0.05 mag in $A_V$ from the Schlegel et
al. values, except for IC3051 and NGC4318 where they are $\sim$ 0.1
mag lower. Uncorrected internal extinction, which would strictly need
to be modelled as embedded rather than as a screen, might lead to ages
being overestimated.

\subsection{Metallicity determination}

\begin{figure*}
  \begin{minipage}[c]{8cm}
    \vspace{0pt}
    \includegraphics[width=8cm]{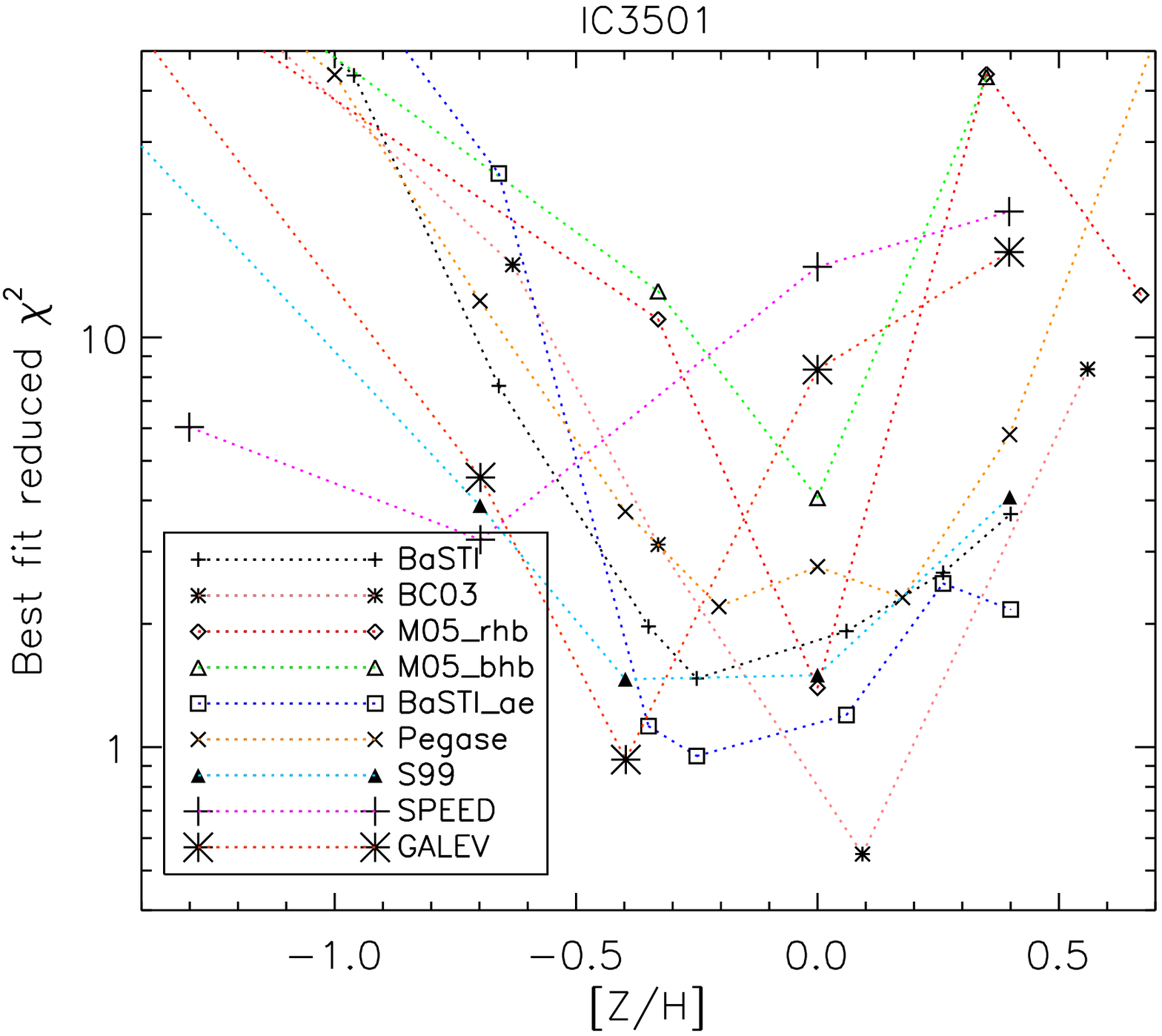}
  \end{minipage}
  \hfill
  \begin{minipage}[c]{8cm}
    \vspace{0pt}
    \includegraphics[width=8cm]{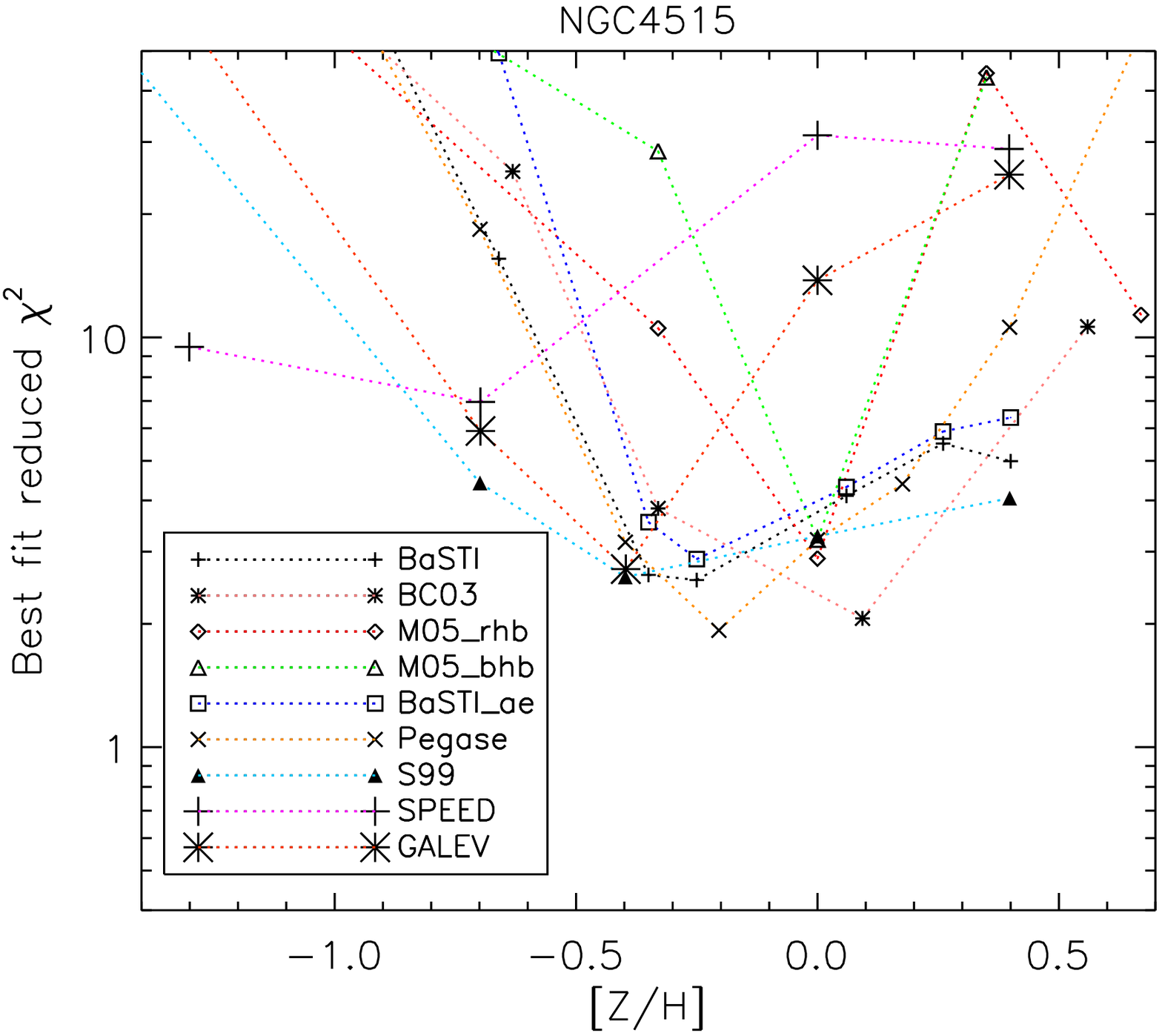}
  \end{minipage}

  \begin{minipage}[c]{8cm}
    \vspace{0pt}
    \includegraphics[width=8cm]{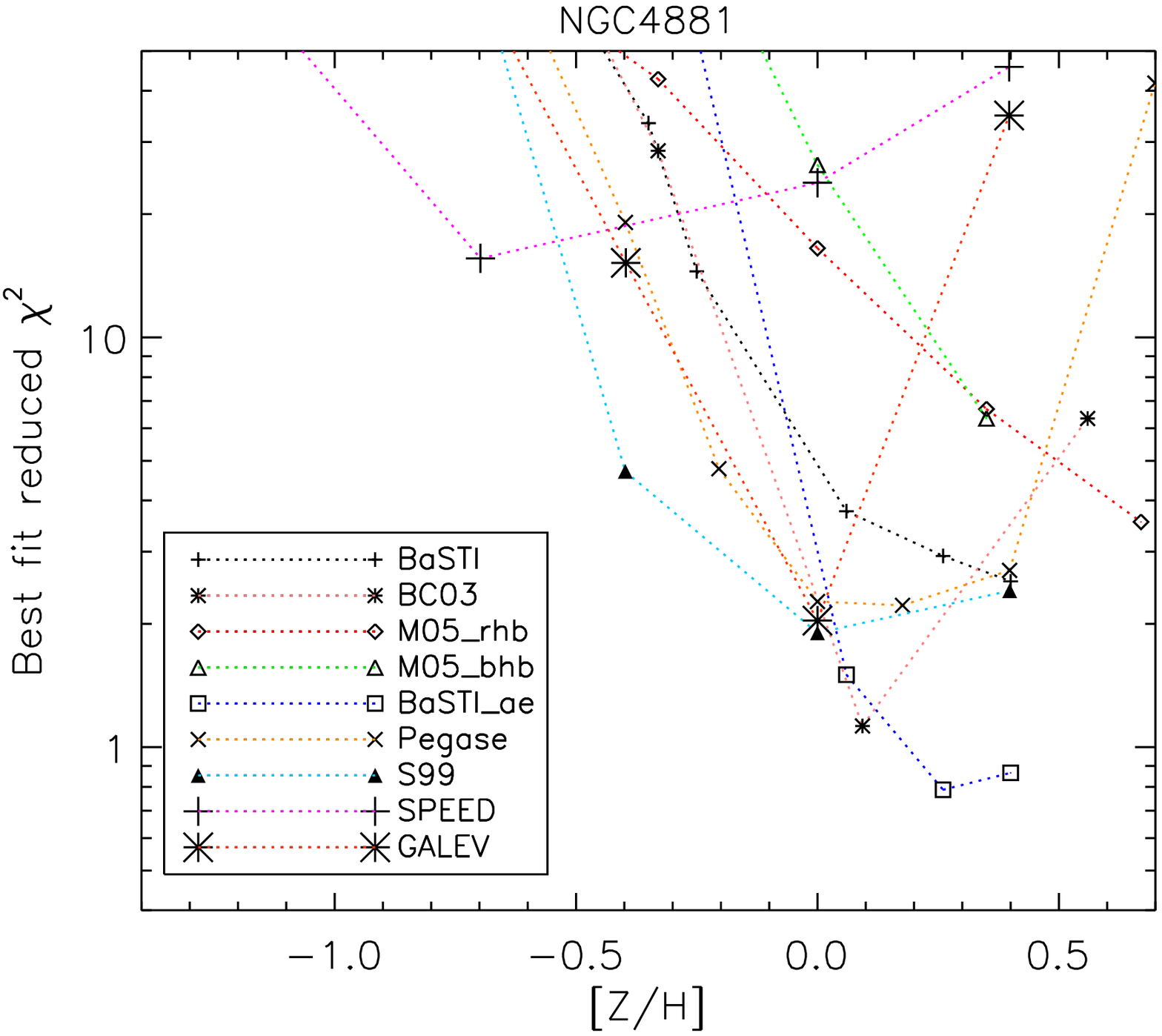}
  \end{minipage}
  \hfill
  \begin{minipage}[c]{8cm}
    \vspace{0pt}
    \includegraphics[width=8cm]{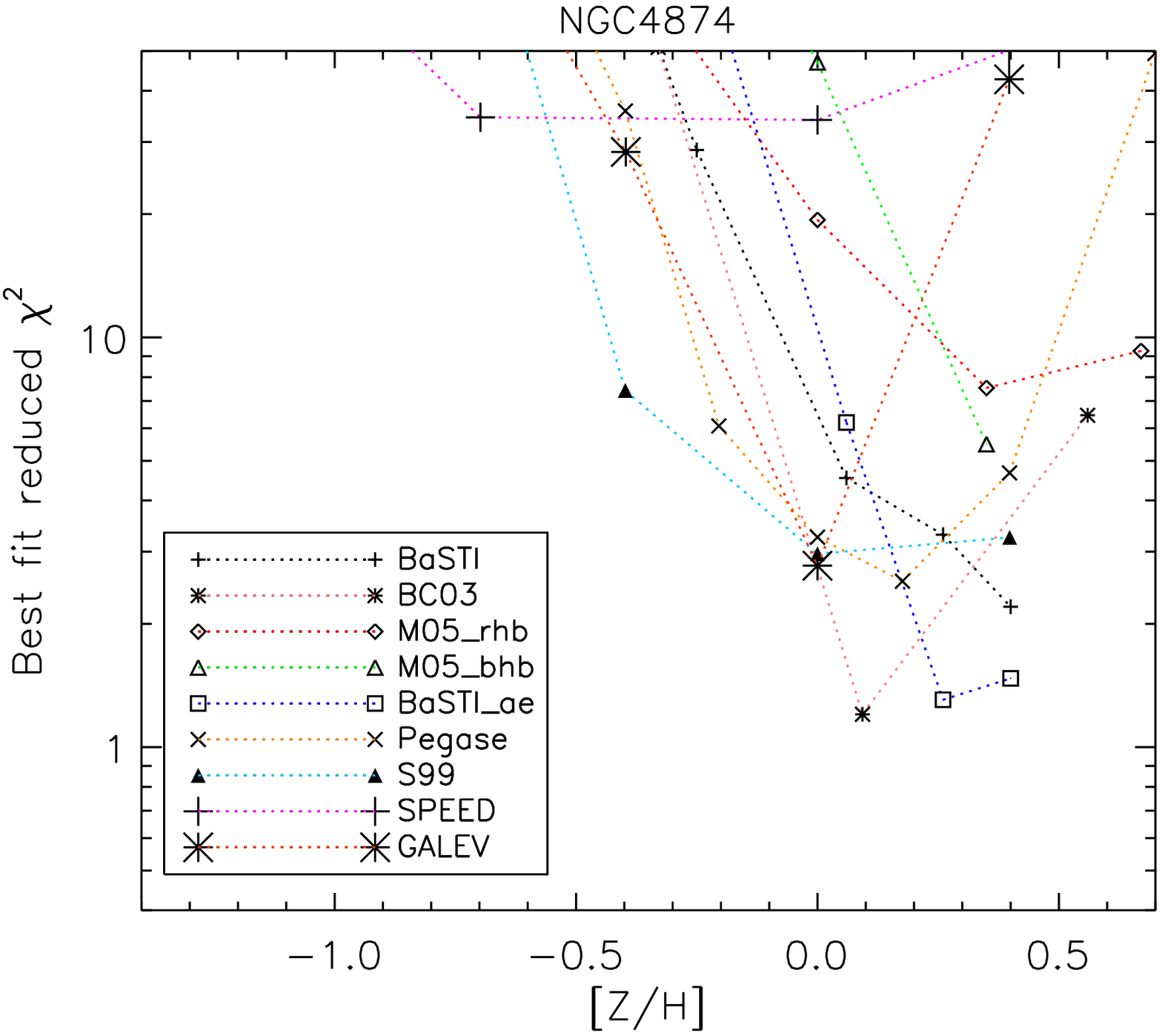}
  \end{minipage}
\caption{$\chi^2_{\nu}$ for best fitting age against [Z/H] for each model
  set, for IC3051 (top left), NGC4515 (top right), NGC4881 (bottom left)
and NGC4874 (bottom right). Models are denoted by the symbols, and different metallicities
connected by the line colours and styles, as denoted in the key in the box.}
\label{fig:metallicity}
\end{figure*}

In Figure ~\ref{fig:metallicity} we plot the lowest value of $\chi^2_{\nu}$,
for any age, against [Z/H] for each model set. For BaSTI
$\alpha$-enhanced models [Z/H] = [Fe/H] + 0.35, for all other models
we assume [Z/H] $\simeq$ [Fe/H].

With the exception of the SPEED and Maraston models there is broad
agreement between the models sets on [Z/H]. However the broad minima
in the $\chi^2_{\nu}$ distributions illustrate further the significant
uncertainties in estimates of [Z/H] or [Fe/H] from broad-band
photometry.

\subsection{Comparison with spectroscopic ages and metallicities}

A useful comparison sample for the present work is provided by Trager
et al. (2008), who present age, metallicity and [$\alpha$/Fe]
estimates from spectroscopic line index measurements for a sample of
12 Coma cluster early-type galaxies, including 5 from our sample.
However, it is important to note that the models used by Trager et
al. (2008) are based on isochrones from Worthey (1994), unlike all of
the models tested in this paper, so differences found may in part be
due to the underlying stellar models.  The quantitative effects
  on model predictions of line indices and spectral energy
  distributions due to changes in models are uncertain.  Effects
  resulting from e.g. differing temperature and metallicity scales are 
being studied by Percival \& Salaris (2009, in preparation).  

In table ~\ref{tab:tragerdata} we reproduce the
age and metallicity measurements from Table 5 of Trager et al.
(2008).  The quantities they tabulate are [Z/H] and [E/Fe], where E
stands for Enhanced elements. [E/Fe] does not precisely correspond to
[$\alpha$/Fe] in the BaSTI models, as the elements enhanced are
somewhat different, as discussed by Trager et al. (2000a). However the
key element for measuring [E/Fe] is Mg, which is an $\alpha$-element.

\begin{table*}
\caption{Comparison with spectroscopic measurements from Trager et al. (2008)}
\label{tab:tragerdata}
\begin{tabular}{lcccccc}
\hline
 &\multicolumn{6}{c}{Trager et al.}\\
 &Age (Gyr)&$\sigma_{Age}$&[Z/H]&$\sigma_{[Z/H]}$&[E/Fe]&$\sigma_{[E/Fe]}$\\ \hline
NGC4867&3.0&0.3&0.54&0.04&0.20&0.01 \\
NGC4872&4.8&0.4&0.36&0.02&0.18&0.01 \\
NGC4871&4.5&0.4&0.36&0.04&0.14&0.01 \\
NGC4873&4.5&0.4&0.32&0.04&0.19&0.01 \\
NGC4874&7.9&1.0&0.38&0.04&0.17&0.01 \\ \hline
\end{tabular}

{\it Notes to Table \ref{tab:tragerdata}:} Where Trager et al. (2008) give asymmetric error bars,
we quote here the larger value.
\end{table*}

Trager et al. [Z/H] values are consistent with [Fe/H] as found from BaSTI
and P\'{E}GASE scaled solar models,  and, using the relation 
[Z/H]=[Fe/H]+0.35 for the BaSTI $\alpha$-enhanced
models, they are consistent with these also. BC03 do not have sufficient metallicity resolution for an
adequate comparison. Trager et al. ages are older than the best fits
for BaSTI and P\'{E}GASE scaled solar models, but younger than those
found for the $\alpha$-enhanced models. This might imply that models with
[$\alpha$/Fe] $\sim$ +0.2 could provide both a better fit to the data and a
better match to Trager et al. (2008), but this is impossible to test
until such models are available.

Trager et al. (2008), and indeed much of the current work on 
stellar populations using spectroscopic data, uses the Lick index system.
This is defined from fairly low resolution spectra ($\sim$8.4~\AA~FWHM), 
however for massive galaxies the effective resolution
of the spectra is limited to this by the internal velocity dispersion
($\sigma \sim$ 200km/s). Ages measured on the Lick system have been shown
to suffer from age-metallicity degeneracy, and also to be somewhat dependent 
upon which Balmer line index is used (e.g. Puzia et al.
2005; Brodie et al. 2005). Alternative index definitions have been proposed
(Vazdekis \& Arimoto 1999; Cervantes \& Vazdekis 2009) which to some extent 
lift this degeneracy. In a future paper, we will compare the full, high-resolution
BaSTI model spectra (Percival et al. 2009), with  higher resolution spectra of 
lower velocity dispersion galaxies (e.g. Smith et al. 2009) to derive single or 
composite stellar population parameters.

\subsection{Comparison with predictions from SBF magnitudes}

 Lee et al. (2009), using BaSTI isochrones, investigate
  predictions for the $I$-band Surface Brightness Fluctuation (SBF)
  magnitudes from scaled solar and $\alpha$-enhanced isochrones. Using
  data from Blakeslee et al. (2009), they find the $\alpha$-enhanced
  isochrones to be the better fit to the redder and more massive
  galaxies, whereas the scaled-solar models work better for bluer and
  less massive galaxies. They attribute this to the effect of oxygen
  enhancement on the upper RGB and AGB. In general this result agrees
  with what we see in our data, for instance in Table
  ~\ref{tab:basticomparison} we see that the $\alpha$-enhanced models
  provide lower values of $\chi^2$ for the most massive galaxies (this
  table is ordered by increasing luminosity). However there is still
  some work required to reconcile models, SBF magnitudes, and
  integrated colours, and indeed Lee et al. show that there are
  significant differences between Padova and BaSTI model predictions
  of ($V-I$) and $I$-band SBF magnitude, for the same scaled-solar
  abundances.

\subsection{Effect of the IMF}

In this study we have examined the effect of the choice of model set
on the derived SSP parameters, however we have not examined different
IMFs, having used the Kroupa (2001) IMF for all but one of our
datasets, nor have we examined in detail the parametrisation of the
treatment of the TP-AGB. Conroy et al. (2009) examined the effect of
the TP-AGB and IMF using a dataset containing a much larger sample,
but at lower photometric precision. It would be instructive to analyse
our data with the Conroy et al. models,  furthermore in a future paper 
we shall investigate, using the BaSTI models, the effect of different 
parametrisations of the IMF upon the derived colours.

\section{Conclusions}

We find that SSP models provide good fits to broad-band photometry
from $u$ to $K_s$ bands, for a variety of morphologically early-type galaxies. 
Although we have tested only SSP models, we can at least say that the 
broad-band data do not require a more complicated star formation history.
For galaxies brighter than $M_V \sim -20.5$ 
the best fits are usually provided by fully self-consistent $\alpha$-enhanced models,
which are currently only available for BaSTI. This is consistent with a scenario in which 
the stars in massive galaxies formed over a relatively short time interval (Thomas et al. 2005).
For galaxies fainter than $M_V \sim -20.5$ 
although the popular Bruzual \& Charlot (2003) models provide in most cases objectively the 
best fit to the data, other scaled-solar  abundance models including P\'{E}GASE, GALEV, 
Starburst99 and BaSTI provide very similar solutions of very similar goodness of fit. This is 
consistent with formation of the stellar content of these galaxies over a longer time interval. We find 
that SPEED and Maraston (2005) models do not provide good fits to the data, and that the best 
fits they do provide occur at unrealistically low metal abundance and unrealistically young 
age respectively.

Broad-band photometry from $u$ to $K_s$, in the presence of realistic photometric errors, 
does not fully break the age-metallicity degeneracy, and the models which have sufficient
metallicity resolution show that there are strongly anticorrelated
uncertainties of $\simeq 0.25$ in log age, and $\simeq 0.18$ in log
metallicity, on the derived population parameters. BaSTI $\alpha$-enhanced
models fit at older ages than the scaled-solar models, but age-metallicity degeneracy is
still present. 

BaSTI models use stellar tracks with two values of both [$\alpha$/Fe]
and the Reimers (1975) mass-loss parameter $\eta$, which controls horizontal
branch morphology. We find that the latter has a much smaller effect than the former, but 
the range of this parameter that we have explored in this paper is small.

To improve our knowledge of stellar population parameters, it is important to investigate 
complementary techniques. These will include extending the wavelength range used further
into the ultraviolet with HST/WFC3, which will provide an additional tool for breaking the age-metallicity
degeneracy. However to utilize the UV photometry it is particularly important to understand the
contribution of the regions of the HR diagram which contribute most to
the UV flux, in particular the Horizontal Branch, Blue Stragglers, and
any contribution from a young stellar component, for instance from
merger induced star formation. Further complementary techniques include 
studies of the properties of star clusters associated with the galaxies, which 
sometimes show subpopulations indicating different stellar population parameters, 
and study of surface brightness fluctuations at near infra-red wavelengths,  which are a powerful
probe of the structure of the AGB and upper RGB.

\section*{Acknowledgments}

We thank St\'{e}phane Charlot for providing data in advance of publication. This work is 
supported by the Science and Technology Facilities Council under rolling grant PP/E001149/1,
``Astrophysics Research at Liverpool John Moores University''.

Funding for the SDSS and SDSS-II has been provided by the Alfred P. Sloan Foundation, the 
Participating Institutions, the National Science Foundation, the U.S. Department of Energy, 
the National Aeronautics and Space Administration, the Japanese Monbukagakusho, the Max 
Planck Society, and the Higher Education Funding Council for England. 
The SDSS Web Site is http://www.sdss.org/.

The SDSS is managed by the Astrophysical Research Consortium for the Participating Institutions. 
The Participating Institutions are the American Museum of Natural History, Astrophysical Institute 
Potsdam, University of Basel, University of Cambridge, Case Western Reserve University, University 
of Chicago, Drexel University, Fermilab, the Institute for Advanced Study, the Japan Participation 
Group, Johns Hopkins University, the Joint Institute for Nuclear Astrophysics, the Kavli Institute 
for Particle Astrophysics and Cosmology, the Korean Scientist Group, the Chinese Academy of Sciences 
(LAMOST), Los Alamos National Laboratory, the Max-Planck-Institute for Astronomy (MPIA), the 
Max-Planck-Institute for Astrophysics (MPA), New Mexico State University, Ohio State University, 
University of Pittsburgh, University of Portsmouth, Princeton University, the United States Naval 
Observatory, and the University of Washington.

This publication makes use of data products from the Two Micron All Sky Survey, which is a joint 
project of the University of Massachusetts and the Infrared Processing and Analysis Center/California 
Institute of Technology, funded by the National Aeronautics and Space Administration and the 
National Science Foundation; and of the NASA/IPAC Extragalactic Database (NED) which is
operated by the Jet Propulsion Laboratory, California Institute of
Technology, under contract with the National Aeronautics and Space
Administration.  

We thank an anonymous referee for valuable comments, and also David Hill of St Andrews University
for posing questions about SDSS zero points which enabled us to eliminate a potential source 
of systematic error.

\label{lastpage}


\begin{thebibliography}{}
\bibitem[Adelman-McCarthy et al. (2008)]{SDSSDR6} Adelman-McCarthy, J.K.
et al. 2008, ApJS, 175, 297
\bibitem[Alongi et al. (1993)]{Alo93} Alongi, M., Bertelli, G., Bressan, A., Chiosi, C., 
Fagotto, F., Greggio, L. \& Nasi, E. 1993, A\&AS, 97, 851
\bibitem[Anders \& Fritze-v. Alvensleben (2003)]{And03} Anders, P. \& Fritze-v. Alvensleben,
U. 2003, A\&A, 401, 1063
\bibitem[Anders et al. (2004)]{And04} Anders, P., Bissantz, N., Fritze-v. Alvensleben, 
U. \& de Grijs, R. 2004, MNRAS, 347, 196
\bibitem[Bell \& de Jong (2000)]{BdJ00} Bell, E.F. \& de Jong, R.S. 2000, MNRAS, 312, 497
\bibitem[Blakeslee et al. (2009)]{Bla09} Blakeslee, J.P. et al. 2009, ApJ, in press (arXiv:0901.1138)
\bibitem[Bothun et al. (1984)]{Both84} Bothun, G.D., Romanishin, W., Strom, S.E. \&
Strom, K.M. 1984, AJ, 89, 1300
\bibitem[Bothun \& Gregg (1990)]{BG90} Bothun, G.D. \& Gregg, M.D. 1990, ApJ, 350, 73 
\bibitem[Bressan et al. (1993)]{Bre93} Bressan, A., Fagotto, F., Bertelli, G. \&
Chiosi, C. 1993, A\&AS, 100, 647
\bibitem[Brodie et al. (2005)]{Bro05} Brodie, J.P., Strader, J., Denicol\'o, G., Beasley, M.A.,
Cenarro, A.J., Larsen, S.S. \& Kuntschner, H. 2005, AJ, 129, 2643
\bibitem[Bruzual \& Charlot (1993)]{BC93} Bruzual, G. \& Charlot, S. 1993,
ApJ, 405, 538 
\bibitem[Bruzual \& Charlot (2003)]{BC03} Bruzual, G. \& Charlot, S. 2003,
MNRAS, 344, 1000
\bibitem[Burstein \& Heiles (1982)]{BH82} Burstein, D. \& Heiles, C. 1982, AJ, 87, 1165
\bibitem[Busso et al. (2007)]{Bus07} Busso, G., et al. 2007, A\&A, 474, 105
\bibitem[Cassisi et al. (1997a)]{Cas97a} Cassisi, S., Castellani, M. \& Castellani, V. 1997a,
A\&A, 317, 108
\bibitem[Cassisi et al. (1997b)]{Cas97b} Cassisi, S., degl'Innocenti, S. \& Salaris, M. 1997b,
MNRAS, 290, 515
\bibitem[Cassisi et al. (2000)]{Cas00} Cassisi, S., Castellani, V., Ciarcelluti, P., 
Piotto, G. \& Zoccali, M. 2000, MNRAS, 315, 679
\bibitem[Castelli \& Kurucz (2003)]{CK03} Castelli, F. \& Kurucz, R.L. 2003,
in ``Modelling of Stellar Atmospheres'', proceedings of the 210th symposium of the IAU,
eds: N. Piskunov, W.W. Weiss, \& D.F. Gray (ASP: San Francisco), pA20
\bibitem[Cervantes \& Vazdekis (2009)]{Cer09} Cervantes, J.L. \& Vazdekis, A. 2009, 
MNRAS, 392, 691
\bibitem[Chabrier (2003)]{Cha03} Chabrier, G. 2003, PASP, 115, 763
\bibitem[Charlot \& Bruzual (1991)]{CB91} Charlot, S. \& Bruzual, G. 1991,
ApJ, 367, 126 
\bibitem[Clayton \& Cardelli (1988)]{CC88} Clayton, G.C. \& Cardelli, J.A. 1988,
AJ, 96, 695
\bibitem[Cohen, Wheaton \& Megeath (2003)]{CWM03} Cohen, M., Wheaton, Wm. A., \& Megeath, S.T., 2003, AJ, 126, 1090
\bibitem[Conroy et al. (2009)]{Con09} Conroy, C., Gunn, J.E. \& White, M. 2009,
ApJ, submitted (arXiv:0809.4261)
\bibitem[Cordier et al. (2007)]{Cor07} Cordier, D., Pietrinferni, A., Cassisi, S., \&
Salaris, M. 2007, AJ, 133, 468
\bibitem[C\^ot\'e et al. (2004)]{Cote04} C\^ot\'e, P., et al.\ 2006, ApJS, 
153, 223
\bibitem[de Grijs et al. (005)]{deG05} de Grijs, R., Anders, P., Lamers, H.J.G.L.M., 
Bastian, N., Fritze-v. Alvensleben, U., Parmentier, G., Sharina, M.E. \& Yi, S.
2005, MNRAS, 359, 874
\bibitem[de Vaucouleurs et al.(1991)]{RC3} 
de Vaucouleurs, G., de Vaucouleurs, A., Corwin, H. G., Buta, R. J.,
Paturel, G., \& Fouque, P. 1991, Third Reference Catalogue of Bright
Galaxies (New York: Springer-Verlag)
\bibitem[Eggleton (1971)]{Egg71} Eggleton, P.P. 1971, MNRAS, 151, 351
\bibitem[Eggleton (1972)]{Egg72} Eggleton, P.P. 1972, MNRAS, 156, 361
\bibitem[Eisenstein et al. (2006)]{Eis06} Eisenstein, D.J. et al. 2006, ApJS,
167, 40.
\bibitem[Eminian et al. (2008)]{Em08} Eminian, C., Kauffmann, G., Charlot, S.,
Wild, V., Bruzual, G., Rettura, A. \& Loveday, J. 2008, MNRAS, 384, 930
\bibitem[Fagotto et al. (1994a)]{Fag94a} Fagotto, F., Bressan, A., Bertelli, G. \&
Chiosi, C. 1994a, A\&AS, 104, 365
\bibitem[Fagotto et al. (1994b)]{Fag94b} Fagotto, F., Bressan, A., Bertelli, G. \&
Chiosi, C. 1994b, A\&AS, 105, 29
\bibitem[Fioc \& Rocca-Volmerange (1997)]{FRV97} Fioc, M. \& Rocca-Volmerange, B. 1997,
A\&A, 326, 950
\bibitem[Fioc \& Rocca-Volmerange (1999)]{FRV99} Fioc, M. \& Rocca-Volmerange, B. 1999,
arXiv:astro-ph/9912179
\bibitem[Girardi et al. (1996)]{Gir96} Girardi, L., Bressan, A., Chiosi, C.,
Bertelli, G. \& Nasi, E. 1996, A\&AS, 117, 113
\bibitem[Girardi et al. (2000)]{Gir00} Girardi, L., Bressan, A., Bertelli, G.
\& Chiosi, C. 2000, A\&AS, 141, 371
\bibitem[Groenewegen \& de Jong (1993)]{GdJ93} Groenewegen, M.A.T. \& de Jong, T. 1993,
A\&A, 267, 410
\bibitem[Hempel \& Kissler-Patig (2004a)]{HKP04a} Hempel, M. \& Kissler-Patig, M. 2004a, 
A\&A, 419, 863
\bibitem[Hempel \& Kissler-Patig (2004b)]{HKP04b} Hempel, M. \& Kissler-Patig, M. 2004b, 
A\&A, 428, 459
\bibitem[Hempel et al. (2005)]{Hem05} Hempel, M., Geisler, D. Hoard, D.W. \& Harris, W.E.
2005, A\&A, 439, 59
\bibitem[James et al. (2006)] {Jam06} James, P.A., Salaris, M., Davies, J.I.,
Phillipps, S. \& Cassisi, S. 2006, MNRAS, 367, 339
\bibitem[Jimenez et al. (1995)]{Jim95} Jimenez, R., Jorgensen, U.G., Thejll, P. \&
Macdonald, J. 1995, MNRAS, 275, 1245
\bibitem[Jimenez et al. (1996)]{Jim96} Jimenez, R., Thejll, P., Jorgensen, U.G.,
Macdonald, J. \& Pagel, B. 1996, MNRAS, 282, 926
\bibitem[Jimenez et al. (2004)]{Jim04} Jimenez, R., Macdonald, J., Dunlop, J., 
Padoan, P. \& Peacock, J.A. 2004, MNRAS, 349, 240
\bibitem[Jorgensen (1991)]{Jor91} Jorgensen, U.G. 1991, A\&A, 246, 118
\bibitem[Kaviraj et al. (2007a)]{Kav07a} Kaviraj, S., Rey, S.-C., Rich, R.M.,
Yoon, S.-J. \& Yi, S.K. 2007a, MNRAS, 381, L74
\bibitem[Kaviraj et al.(2007b)]{Kav07b} Kaviraj, S. et al. 2007b, ApJS, 
173, 619
\bibitem[Kim \& Lee (2007)]{KL07} Kim, S.S. \& Lee, M.G. 2007, PASP, 119, 1449
\bibitem[Kodama \& Arimoto (1997)]{KA97} Kodama, T. \& Arimoto, N. 1997,
A\&A, 320, 41
\bibitem[Koleva et al. (2008)]{Kol08} Koleva, M., Prugniel, Ph., Ocvirk, P.,
Le Borgne, D. \& Soubiran, C. 2008, MNRAS, 385, 1998
\bibitem[Kroupa (2001)]{Kr01} Kroupa, P. 2001, MNRAS, 322, 231
\bibitem[Kundu et al. (2005]{Kun05} Kundu, A. et al. 2005, ApJL, 634, L41
\bibitem[Kuntschner et al. (2001)]{Kun01} Kuntschner, H., Lucey, J.R., Smith, R.J.,
Hudson, M.J. \& Davies, R.L. 2001, MNRAS, 323, 615
\bibitem[Kurucz (1993)]{Kur93} Kurucz, R.L. 1993, ``ATLAS9 Stellar Atmosphere Programs 
and 2 km/s grid, Smithsonian Astrophysical Observatory.'',  CD-ROM No. 13
\bibitem[Lan\c{c}on \& Mouhcine (2002)]{LM02} Lan\c{c}on, A. \& Mouhcine, M. 2002,
A\&A, 393, 167
\bibitem[Le Borgne et al. (2003)] {Leb03} Le Borgne, J.-F. et al. 2003, A\&A, 402, 433
\bibitem[Lee et al. (2007)]{Lee07} Lee, H.-C., Worthey, G., Trager, S.C. \&
Faber, S.M. 2007, ApJ, 664, 215
\bibitem[Lee et al. (2009)]{Lee09} Lee, H.-C., Worthey, G. \& Blakeslee, J.P.,
ApJ, in press (arXiv:0902.1177)
\bibitem[Lejeune (1997)]{Lej97} Lejeune, T., Cuisinier, F \& Buser, R. 1997,
A\&AS, 125, 229
\bibitem[Lejeune (1998)]{Lej98} Lejeune, T., Cuisinier, F \& Buser, R. 1998,
A\&AS, 130, 65
\bibitem[Maraston (1998)]{Mar98} Maraston, C. 1998, MNRAS, 300, 872
\bibitem[Maraston (2005)]{Mar05} Maraston, C. 2005, MNRAS, 362, 799
\bibitem[Maraston et al. (2008)] {Mar08} Maraston, C., Str\"{o}mb\"{a}ck, G.,
Thomas, D., Wake, D.A. \& Nichol, R.C. 2008, ApJL, submitted (arXiv:0809.1867v1)
\bibitem[Marigo \& Girardi (2007)]{MG07} Marigo, P. \& Girardi, L. 2007, A\&A,
469, 239
\bibitem[Michard (2005)]{Mich05} Michard, R. 2005, A\&A, 429, 819
\bibitem[Nelan et al. (2005)]{Nel05} Nelan, J.E., Smith, R.J., Hudson, M.J., Wegner, G.A.,
Lucey, J.R., Moore, S.A.W., Quinney, S.J. \& Suntzeff, N.B. 2005, ApJ, 632, 137.
\bibitem[Ocvirk et al. (2006a)]{Ocv06a} Ocvirk, P., Pichon, C., Lan\c{c}on, A. \&
Thi\'ebaut, E. 2006a, MNRAS, 365, 46
\bibitem[Ocvirk et al. (2006b)]{Ocv06b} Ocvirk, P., Pichon, C., Lan\c{c}on, A. \&
Thi\'ebaut, E. 2006b, MNRAS, 365, 74
\bibitem[Padmanabhan et al. (2008)]{Pad08} Padmanabhan, N. et al. 2008, ApJ, 674, 1217
\bibitem[Panter et al. (2003)]{Pan03} Panter, B., Heavens, A.F. \& Jimenez, R. 2003,
MNRAS, 343, 1145
\bibitem[Panter et al. (2007)]{Pan07} Panter, B., Jimenez, R., Heavens, A.F.
\& Charlot, S. 2007, MNRAS, 378, 1550
\bibitem[Peletier \& Balcells (1996)]{PB96} Peletier, R.F. \& Balcells, M. 1996, AJ, 111, 2238
\bibitem[Peletier et al. (1990)] {Pel90} Peletier, R.F., Valentijn, E.A. \& Jameson, R.F. 1990,
A\&A, 233, 62
\bibitem[Percival et al. (2008)]{Per08} Percival, S.M., Salaris, M., Cassisi, S. \&
Pietrinferni, A. 2008, ApJ, 690, 427  
\bibitem[Pessev et al. (2008)]{Pes08} Pessev, P.M., Goudfrooij, P., Puzia, T.H.
\& Chandar, R. 2008, MNRAS, 385, 1535
\bibitem[Pickles (1998)]{Pick98} Pickles, A.J. 1998, PASP, 110, 863
\bibitem[Pietrinferni et al. (2004)]{Piet04} Pietrinferni, A., Cassisi, S., Salaris, M. \&
Castelli, F. 2004, ApJ, 612, 168
\bibitem[Pietrinferni et al. (2006)]{Piet06} Pietrinferni, A., Cassisi, S., Salaris, M. \&
Castelli, F. 2006, ApJ, 642, 797
\bibitem[Poggianti et al. (2001a)]{Pog01a} Poggianti, B.M. et al. 2001a,
ApJ, 562, 689
\bibitem[Poggianti et al. (2001b)]{Pog01b} Poggianti, B.M. et al. 2001b,
ApJ, 563, 118
\bibitem[Puzia et al. (2002)]{Puz02} Puzia, T.H., Zepf, S.E., Kissler-Patig, M.,
Hilker, M., Minniti, D. \& Goudfrooij, P. 2002, A\&A, 391, 453
\bibitem[Puzia et al. (2005)]{Puz05} Puzia, T.H., Kissler-Patig, M., Thomas, D.,
Maraston, C., Saglia, R.P., Bender, R., Goudfrooij, P. \& Hempel, M. 2005, A\&A, 439, 997
\bibitem[Rakos et al. (2007)]{Rak07} Rakos, K., Schombert, J. \& Odell, A. 2007,
ApJ, 658, 929
\bibitem[Rakos et al. (2008)]{Rak08} Rakos, K., Schombert, J. \& Odell, A. 2008,
ApJ, 677, 1019
\bibitem[Rawle et al. (2008)]{Raw08} Rawle, T.D., Smith, R.J., Lucey, J.R., Hudson, M.J.
\& Wegner, G.A. 2008, MNRAS, 385, 2097
\bibitem[Reimers (1975)]{Rei75} Reimers, D. 1975, Mem. Soc. R. Sci. Li\`ege, 6, 369
\bibitem[Renzini \& Buzzoni (1986)]{RB86} Renzini, A. \& Buzzoni, A. 1986,
in ``Spectral evolution of galaxies'', eds: G. Chiosi \& A. Renzini (D. Reidel: Dordrecht),
p195
\bibitem[S\'anchez-Bl\'azquez et al. (2006)]{S-B06} S\'anchez-Bl\'azquez, P., Gorgas, J.,
Cardiel, N. \& Gonz\'alez, J.J. 2006, A\&A, 457, 809
\bibitem[Schlegel et al. (1998)]{Schlegel98} Schlegel, D.J., Finkbeiner, D.P. \& Davis, M.
1998, ApJ, 500, 525 
\bibitem[Schulz et al. (2002)] {Schu02} Schulz, J., Fritze-v. Alvensleben, U., 
M\"{o}ller, C.S. \& Fricke, K.J. 2002, A\&A, 392, 1
\bibitem[Smith et al. (2009)]{Smi09} Smith, R.J., Lucey, J.R., Hudson, M.J., Allanson, S.P., 
Bridges, T.J., Hornschemeier, A.E., Marzke, R.O. \& Miller, N.A. 2009, MNRAS, 392, 1265
\bibitem[Skrutskie et al. (2006)]{2MASS} Skrutskie, M.F. et al. 2006,
AJ, 131, 1163
\bibitem[Thomas et al. (2005)] {Tho05} Thomas, D., Maraston, C., Bender, R. \& Mendes
de Oliveira, C. 2005, ApJ, 621, 673.
\bibitem[Tojeiro et al. (2007)]{Toj07} Tojeiro, R., Heavens, A.F., Jimenez, R. \&
Panter, B. 2007, MNRAS, 381, 1252
\bibitem[Trager et al. (2000a)]{Trager00a} Trager, S.C., Faber, S.M., Worthey, G,
Gonz\'alez, J.J. 2000a, AJ 119, 1645.
\bibitem[Trager et al. (2000b)]{Trager00b} Trager, S.C., Faber, S.M., Worthey, G,
Gonz\'alez, J.J. 2000b, AJ 120, 165.
\bibitem[Trager et al. (2008)]{Trager08} Trager, S.C., Faber, S.M. \& Dressler, A.
2008, MNRAS, 386, 715
\bibitem[Vassiliadis \& Wood (1993)]{VW93} Vassiliadis, E. \& Wood, P.R. 1993, ApJ,
413, 641
\bibitem[Vassiliadis \& Wood (1994)]{VW94} Vassiliadis, E. \& Wood, P.R. 1994, ApJS,
92, 125
\bibitem[Vazdekis \& Arimoto (1999)]{VA99} Vazdekis, A. \& Arimoto, N. 1999,
ApJ, 525, 144
\bibitem[V\'azquez \& Leitherer (2005)]{VL05} V\'azquez, G.A. \& Leitherer, C. 2005,
ApJ, 621, 695 
\bibitem[Westera et al. (2002)]{Wes02} Westera, P., Lejeune, T., Buser, R., Cuisinier, F. \&
Bruzual, G. 2002, A\&A, 381, 524
\bibitem[Worthey 91994)]{Wor94} Worthey, G. 1994, ApJS, 95, 107
\end{thebibliography}
\end{document}